\documentclass[aps,prd,showpacs,notitlepage,nofootinbib,superscriptaddress,floatfix,showkeys,twocolumn]{revtex4-1}
\pdfoutput=1
\usepackage{graphicx}
\usepackage{amsmath,calligra,mathrsfs}
\usepackage{amsfonts}
\usepackage{amssymb}
\usepackage{physics}
\usepackage{tabularx}
\usepackage{multirow}
\usepackage{longtable}
\usepackage{array}
\usepackage{float}
\usepackage{xcolor, soul}
\usepackage{epstopdf}
\usepackage{graphicx}
\usepackage{amsmath}
\usepackage{amsfonts}
\usepackage{amssymb}
\usepackage{epstopdf}
\usepackage{float}
\usepackage{subfigure}
\usepackage{float}
\usepackage{subfigure}
\usepackage{hyperref}
\usepackage{ulem}
\hypersetup{
     colorlinks   = true,
     citecolor    = red,
     linkcolor    = blue,
     urlcolor     = blue,
}\DeclareGraphicsExtensions{.eps} 
\def\beq{\begin{equation}}
\def\eeq{\end{equation}}
\def\bea{\begin{eqnarray}}
\def\eea{\end{eqnarray}}
\def\be{\begin{equation}}
\def\ee{\end{equation}}

\def\bse{\begin{subequations}}
\def\ese{\end{subequations}}

\def\aend{a_{\rm end}}

\def\nuth{\nu_{\rm R}^3}

\def\are{a_{\rm re}}
\def\aend{a_{\rm end}}
\def\tre{T_{\rm re}}
\def\kre{k_{\rm re}}
\def\kend{k_{\rm end}}

\def\mp{M_{\rm p}}
\def\wphi{w_{\rm \phi}}
\def\hend{H_{\rm end}}
\def\lmd{\lambda_{\rm hs}}
\def\ms{m_{\rm s}}
\def\mh{m_{\rm h}}
\def\tqc{T_{\rm qcd}}
\def\tos{T_{\rm osc}}
\def\ak{\mathfrak a}
\def\ma{m_{\rm\ak}}
\def\mat{\tilde{m}_{\rm\ak}}
\def\rha{\rho_{\rm\ak}}

\graphicspath{{./figs/}}
\begin{document}
\newcommand{\red}{\color{red}}
\newcommand{\blue}{\color{blue}}
\newcommand{\green}{\color{green}}
\newcommand{\magenta}{\color{magenta}}
\newcommand{\purple}{\color{purple}}
\newcommand{\mi}{\mathrm{i}}
\newcommand{\rbh}{\rho_{\rm BH}}
\newcommand{\rphi}{\rho_{\phi}}
\title{Thermal and nonthermal dark matters with gravitational neutrino reheating}

\author{Md Riajul Haque}%
\email{riaj.0009@gmail.com}
\affiliation{Centre for Strings, Gravitation and Cosmology,
Department of Physics, Indian Institute of Technology Madras, 
Chennai~600036, India}
\author{Debaprasad Maity}
\email{debu@iitg.ac.in}
\author{Rajesh Mondal}%
\email{mrajesh@iitg.ac.in}
\affiliation{%
	Department of Physics, Indian Institute of Technology Guwahati, Guwahati 781039, Assam, India}%

\date{\today}

\begin{abstract}
We have discussed in detail how neutrinos produced from inflaton solely through gravitational interaction can successfully reheat the universe. For this, we have introduced the well-known Type-I seesaw neutrino model. Depending on seesaw model parameters, two distinct reheating histories have been realized and dubbed as 
i) Neutrino dominating: Following the inflaton domination, the universe becomes neutrino dominated, and their subsequent decay concludes the reheating process, and ii) Neutrino heating: Despite being sub-dominant compared to inflaton energy, neutrinos efficiently heat the thermal bath and produce the radiation dominated universe.
Imposing baryon asymmetric yield, the $\Delta N_{\rm eff}$ constraint at Big Bang Nucleosynthesis (BBN) considering primordial gravitational waves (PGW), we have arrived at the following constraints on reheating equation of state to lie within $0.5\lesssim w_\phi\lesssim1.0$. In these neutrino-driven reheating backgrounds, we further performed a detailed analysis of both thermal and non-thermal production of dark matter (DM), invoking two minimal models, namely the Higgs portal DM and classical QCD pseudo scalar axion. An interesting correlation between seemingly uncorrelated DM and Type-I seesaw parameters has emerged when confronting various direct and indirect observations. When DMs are set to freeze-in, freeze-out, or oscillate during reheating, new parameter spaces open, which could be potentially detectable in future experiments, paving an indirect way to look into the early universe in the laboratory.     
\end{abstract}
\maketitle
\section{Introduction}
Inflatons are the dominant energy component right after the termination of inflation \cite{r4,r5,r6}. The current observable universe however, demands a mechanism of transferring inflaton energy into standard model particles dubbed as reheating \cite{r1,r2,r3,b1,Kofman:1994rk}. If the decay of inflaton is primarily dominated by gravitational interactions, can the universe be populated by radiation? Many investigations in the recent past  \cite{Haque:2022kez,Clery:2021bwz,Clery:2022wib,Dimopoulos:2018wfg,Nakama:2018gll,Haque:2023yra,Kunimitsu:2012xx,Hashiba:2018iff} indicate that the answer to this question is in the affirmative: There does exist a
scenarios which require detailed 
exploration, and recently proposed gravitational reheating (GRe) \cite{Haque:2022kez} is one such interesting example. Particularly due to its lack of new physics requirements in the inflaton sector and, of course, robust prediction, such a scenario warrants intense scrutiny on its consistency with different observations. Two indirect observations related to the minimum reheating temperature set by Big-Bang Nucleosynthesis (BBN), $T_{\rm re}^{\rm min} \sim T_{\rm BBN} \simeq 4$ MeV \cite{Kawasaki:2000en,Sarkar:1995dd,Hannestad:2004px}, and the constraints \cite{Mishra:2021wkm,Haque:2021dha} on the total GW energy density $\Omega^{\rm k}_{\rm GW} h^2 \sim 1.7\times 10^{-6}$ \cite{Pagano:2015hma,Yeh:2022heq,Planck:2018vyg} from the contribution to the extra relativistic degrees of freedom at the time of BBN already restricts GRe model to be strictly kination dominated\cite{Haque:2022kez}.

Recently PBH reheating has been proposed to relax such constraint, which renders the scenario viable for any inflaton equation of state $\wphi > 1/3$ \cite{RiajulHaque:2023cqe,Haque:2023lzl}. In spite of its attractive nature and recent interest in the community, the PBH formation mechanism itself is still under active investigation, let alone its observational prospects, which are still not confirmed. 

Keeping all these aspects into consideration, we recently proposed a gravitational reheating scenario \cite{Haque:2023zhb} governed by right-handed neutrinos in the well-known Type-I seesaw framework embedded in the standard model (SM). We call such a reheating scenario gravitational neutrino reheating and in short $\nu$GRe. Crucial aspects of such consideration are that apart from successfully explaining the observed light neutrino masses and baryogenesis, the model can successfully reheat the universe via purely gravitational interaction, satisfying all the aforementioned constraints \cite{Haque:2023zhb}. 

 In this work, we will explore this scenario in greater detail and examine its impact on various dark matter (DM) models. 
Phenomenological studies of DM were mostly confined in the early radiation-dominated universe \cite{fo1,fo2,fo3,fo4,fo5,fo6,fo7,fo8} with severe constraints from observation. 
Further, the direct probe of the reheating phase is experimentally challenging. Studying impact of the reheating phase on DM phenomenology, therefore, has gain widespread interest in the recent past \cite{Maity:2018dgy,Garcia:2020eof,Garcia:2020wiy,Garcia:2021gsy,Maity:2018exj,Haque:2020zco,Giudice:2000ex,Haque:2021mab,Barman:2022tzk,Bhattiprolu:2022sdd,Harigaya:2014waa,Harigaya:2019tzu,Okada:2021uqk,Ghosh:2022fws,Ahmed:2022tfm,Bernal:2022wck,Bernal:2023ura,Haque:2023yra}. 
Non-standard reheating histories have been shown to allow previously unexplored DM parameter regions and, therefore, opened up the intriguing possibilities of putting indirect constraints on the early universe phases, such as reheating and inflation via DM experiments.
This further led to significant interest in building up a unified framework of inflaton and DM in conjunction with the reheating phase.  
Gravitational particle production, particularly during inflation and reheating, is universal. Hence, any phenomenological study cannot ignore such production, and that may have a significant impact on the parameter estimation of any DM model under consideration. It is in this parlance that we study the impact of gravitational neutrino reheating ($\nu$GRe) on two popular minimal DM models, namely the Higgs portal DM and QCD Axion. We study in detail both thermal and non-thermal production of DM during reheating, which is governed by gravitational neutrino production and their subsequent decay. We show how such consideration leads to an interesting correlation between the DM and Type-I seesaw parameters confronting direct observations. 

The paper is organized as follows: In Sec.\ref{sc2}, we provide a detailed discussion of the gravitational neutrino reheating framework and identify the corresponding parameter spaces. In Sec.\ref{sc3}, we show how constraints from extra relativistic degrees of freedom during BBN, considering PGWs, put restrictions on $\nu$GRe. In Sec.\ref{sc4}, we discuss the impact of $\nu$GRe on the inflationary parameters ($n_{\rm s}\,,r$).In Sec.\ref{sc5}, we analyze the possible constraints of leptogenesis. In Sec.\ref{sc6} and \ref{sc7}, we consider various possible scenarios corresponding to
 DM production and discuss their possible constraints from
 the perspective of various theoretical and experimental bounds. Finally, conclude in Sec.\ref{sc8}.

\section{Gravitational Neutrino Reheating}{\label{sc2}}
\subsection{The framework}
\begin{figure}[t]
\centering
\includegraphics[width=6cm]{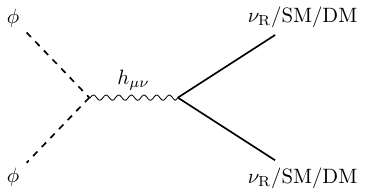}
\caption{Feynman diagram for the production of RHN/SM/DM
through the gravitational scattering of the inflaton condensate.}
\label{feymangra}
\end{figure}

In this section, we discuss the neutrino reheating scenario in more detail and identify the full parameter space. Type-I seesaw \cite{Gell-Mann:1979vob,see1,see2} is the simplest extension of the SM, which can generate active neutrino mass along with its potential application to resolve the matter anti-matter asymmetry. The extension contains three additional right-handed SM singlet massive Majorana neutrinos $\nu_R^i(i=1,2,3)$ with the following lepton number violating Lagrangian 
\begin{equation}
    \mathcal{L}={\cal L}_{\phi} + {\cal L}_{\rm SM} -\frac{1}{2}M_{\rm i}\overline{\nu^{\rm c i}_{\rm R}}\nu^{\rm i}_{\rm R}-y_{ij}\bar L_{\rm i}\Tilde{H}{\nu}^{\rm j}_{\rm R}+h.c.,
\end{equation}
where $\tilde {H} = i\sigma_2 H^*$ and $H (L)$, is the SM Higgs (lepton) doublet. The first two terms are the Lagrangian for inflaton $(\phi)$ and SM, respectively. Therefore, except for usual inflaton parameters, Yukawa coupling matrix $y_{ij}$ and Majorana mass $M=\mbox{diag}(M_1, M_2, M_3)$ are the only ones beyond the SM parameters in the theory. The inflaton to satisfy the approximate shift symmetry, its couplings with any other fields are assumed to be negligibly small. After the termination of inflation, all the particles will be produced quantum mechanically via gravitational interaction from the inflaton condensate (see Fig. (\ref{feymangra})), and they will subsequently be diluted due to expansion. Since the gravitational production rate strictly depends on the inflaton energy density, the dominating contribution in such production is at the beginning of reheating. The relevant Boltzmann equations for different density components for these processes would then be,
\begin{subequations}{\label{}}
\begin{align}
& \dot{\rho_{\rm\phi}}+3H(1+w_\phi)\rho_{\rm\phi}+\Gamma^{T}_{\phi}(1+w_\phi)\rho_{\rm\phi}=0 ,\label{rhophi}\\
&\dot{\rho}_{\rm r}+4H\rho_{\rm r}-\Gamma_{\rm\phi\phi\rightarrow hh}(1+w_{\rm\phi})\rho_{\rm\phi}-\langle E_i\rangle\,\Gamma_{i}\, n_{i}=0\,,\label{rhor} \\
&\dot{n}_{i}+3Hn_{i}-R^{i}_{\rm\phi}+\Gamma_{\rm i}\,n_{i}=0\,, \label{rhoN3}
\end{align}
\end{subequations}
where $\rho_{\rm \phi}$, $\rho_{\rm r}$ are the energy density of the inflaton, radiation respectively. $n_{i}$ is the number density of the Right-handed neutrinos (RHNs). $R^{i}_{\phi}$ is the gravitational production rate associated with the RHNs. Note that gravitational production of RHNs mostly occurs at the beginning of the reheating. Hence, its momentum redshifts as $p = p_{in}/a$ starting from its initial momentum $p_{in}\sim m^{\rm end}_{\rm\phi}$ (inflaton mass at the end of inflation). Where $a$ is the cosmological scale factor. The average energy of the RHNs is, therefore, defined as $\langle E_{\rm i}\rangle= \sqrt{p_i^2+M^2_{i}}$. $\Gamma^{\rm T}_{\rm\phi}$ is the total inflaton decay width considering $\Gamma_{\rm\phi\phi\rightarrow hh}$, the gravitational decay width for the SM Higgs production from scattering of inflaton, and the decay width associated with the production of RHN gravitationally. $\Gamma_{i}$'s are the decay width of the RHNs $\Gamma_{i}={(y^\dagger y)_{\rm ii}}M_{i}/(8\pi)$.

Type-I seesaw model considers three copies of RHNs, two RHNs (namely $\nu^1_{\rm R},\nu^2_{\rm R}$) are sufficient to explain both the light neutrino mass and baryon asymmetry of the universe through their out-of-equilibrium non-thermal decay \cite{asaka,Co:2022bgh,Campbell:1992hd,Kaneta:2019yjn,Bernal:2021kaj}. Therefore, the third RHN $(\nu^3_R)$ stands out, and we consider it to be long-lived such that its decay rate is
 suppressed compared to the other two RHNs.  Therefore, out of all the parameters, the following three parameters
\begin{eqnarray}
    \{\beta = \sqrt{(y^\dagger y)_{\rm 33}}, M_3, w_\phi\},
\end{eqnarray}
related to $\nu^3_R$ and inflaton will be shown to control the reheating dynamics. The remaining parameters will give rise to active neutrino mass and baryogenesis.

We show that it is the weak decay width, $\Gamma_{3} \simeq \beta^2 M_3/(8\pi) \ll \Gamma_1,\Gamma_2$ of $\nu_R^3$, with $\beta^2\simeq(m_1\,M_3/v^2)$ compared to the other two RHNs (see Appendix-\ref{CIP}), that plays crucial role for successful reheating to occur in present scenario. Since $\beta^2 \propto m_1$\footnote{Since we use the normal hierarchy of active neutrino masses here, $\beta^2 \propto m_1$. However, $\nu$GRe is also possible if one considers the inverted hierarchy, in which case $\beta^2 \propto m_3$. See Appendix-\ref{CIP}, where we construct the Yukawa coupling using both hierarchies.}, the non-vanishing $\Gamma_3$ immediately sets the mass of the lightest active neutrino mass $m_1$. In other words, to have a neutrino to control the reheating process along with the successful baryogenesis, we indeed require the lightest active neutrino mass to be non-vanishing. This shows a clear correlation between the neutrino parameter and the reheating parameter.

\subsection{Production of RHN$\mbox{s}$ from inflaton}
All the RHNs are assumed to be unstable in terms of their respective decay width. In the absence of decay, the evolution of RHN number density ($n_{\rm i}$) is expected to be governed by the Boltzmann equation,
\be
\dot{n}_{\rm i}+3Hn_{\rm i}=R^{\rm i}_{\rm\phi} = \frac{\rho^2_{\rm \phi}}{4\pi\gamma^2 M^4_{\rm p}}\frac{M^2_{i}}{m^2_{\rm\phi}}\Sigma_{\rm i}\,, \label{numN}
\ee
where, $R^{\rm i}_{\rm\phi}$ the production rate for $\nu^{\rm i}_{\rm R}$ from inflaton scattering mediated by gravity \cite{Clery:2021bwz,Barman:2021ugy} and
\begin{equation} \label{nurate}
\begin{aligned}
   &\Sigma_{\rm i}= \sum^{\infty}_{\nu=1}\frac{1}{\nu^2}\lvert\mathcal P^{2n}_\nu \rvert^2\left(1-\frac{4M^2_{\rm i}}{\nu^2\gamma^2m^2_{\rm\phi}}\right)^{3/2}\,,\\
   \end{aligned}
\end{equation}
accounts for the sum over the Fourier modes of the inflaton potential, and its numerical values are given in Table-\ref{fouriersum}. While evaluating the interaction rate, the inflaton is treated as a time-dependent external and classical background field which undergoes coherent oscillation, and we parametrize the field as
\begin{equation}
    \phi(t)=\phi_0(t).\mathcal{P}(t)=\phi_0(t)\sum_\nu \mathcal{P}_\nu e^{i\, \nu\, \Omega\, t}\,,
\end{equation}
where  $\phi_0(t)$ represents the decaying amplitude of the oscillation and $\mathcal{P}(t)$ encoding the oscillation of the inflaton with the fundamental frequency calculated to be
\cite{Garcia:2020wiy},
\begin{equation} \label{fre}
\Omega = m_\phi(t)\,\gamma\,,~~~\mbox{where}~~\gamma=\sqrt{\frac{\pi\,n}{2\,n-1}}\frac{\Gamma\left(\frac{1}{2}+\frac{1}{2\,n}\right)}{\Gamma\left(\frac{1}{2\,n}\right)}\,.
\end{equation}
\begin{table}[t]
\caption{Numerical values of the Fourier coefficients:}\label{fouriersum}
\centering
 \begin{tabular}{||c | c |c |c |c||} 
 \hline
 $n(w_\phi)$ & $\sum \frac{1}{\nu^2}\lvert\mathcal P^{2 n}_\nu\rvert^2$ & $\sum \nu\lvert \mathcal{P}^{2 n}_\nu\rvert^2$ & $\sum\lvert \mathcal{P}^{2 n}_\nu\rvert^2$\\ [0.5ex] 
 \hline\hline
 2 (1/3) & 0.014  & 0.141 & 0.063\\
 3 (0.50) & 0.011  & 0.145 & 0.056\\
 4 (0.60)& 0.008 & 0.148 & 0.049 \\
 10 (0.82) & 0.002 & 0.149 & 0.027\\ [1ex] 
 \hline
 \end{tabular}
\end{table}
For our study, as a reference inflation model, we consider the $\alpha$-attractor $E$- model \cite{Kallosh:2013hoa,Ferrara:2014cca,Ueno:2016dim}.
 \bea
\label{pot1}
V(\phi)=\Lambda^4\left(1-e^{-\sqrt{\frac{2}{3\alpha}}\frac{\phi}{M_p}}\right)^{2n}\,,
\eea
where $\Lambda$ is the mass-scale fixed by the CMB power spectrum, which is typically of the order $8\times 10^{15}$ GeV, and the parameter $(\alpha,n) $ controls the shape of the potential. The inflaton equation of states $w_\phi$ is directly related with $n$, via the relation $\wphi=(n-1)/{(n-1)}$ \cite{Garcia:2020eof}.
For this model, the time-dependent inflaton mass can be expressed as 
\cite{Garcia:2020eof,btm1}
\begin{eqnarray} \label{mphi}
m_\phi^2\simeq \frac{2n(2n-1)\Lambda^4}{\alpha_1^{2n} M_{\rm p}^2}\left[\frac{\phi_0}{M_{\rm p}}\right]^{2n-2} 
\sim {(m_\phi^{\rm end})}^2\left(\frac{a}{a_{\rm end}}\right)^{-6\,w_\phi},
\end{eqnarray}
where $\alpha_1=\sqrt{3\alpha/2}$ and $m_\phi^{\rm end}$ is the inflaton mass defined at the end of the inflation
\be\label{massend}
m_\phi^{\rm end}\simeq  \frac{\sqrt{2n\,\left(2n-1\right)}}{\alpha_1}\frac{\Lambda^{\frac{2}{n}}}{M_{\rm p}}\left(\rho_\phi^{\rm end}\right)^{\frac{n-1}{2\,n}}\,.
\ee
We choose $\alpha=1$ throughout our study. During the initial period of reheating, inflaton is the dominant energy component, and neglecting the inflaton decay rate compared to the Hubble rate $H$ from Eq.(\ref{rhophi}) $\rho_{\rm\phi}$ evolves as, 
\begin{eqnarray}\label{rhophis}
    \rho_\phi(a)=\rho^{\rm end}_{\rm\phi}\left(\frac{a}{a_{\rm end}}\right)^{-3(1+w_\phi)}\,,
\end{eqnarray}
where $\rho^{end}_{\rm\phi}=3M_{\rm p}^2H_{\rm end}^2$ is the inflation energy density at the end of inflation. Here, $\hend$ is the Hubble parameter, calculated at the end of the inflation, and $\mp=2.4\times10^{18}$ GeV is the reduced Planck
mass. Using Eqs.  \ref{mphi} and \ref{rhophis} in Eq. \ref{numN} one obtains, 
\begin{equation}\label{Nis}
\begin{aligned}
    n_{\rm i}(a)\simeq \frac{3H^3_{\rm end}M^2_{\rm i}\Sigma_{\rm i}}{2\pi(1-w_\phi)(\gamma m^{\rm end}_\phi)^2} \left(\frac{a}{\aend}\right)^{-3} .\\
    \end{aligned}
\end{equation}
Where the effective initial number density of right-handed neutrinos at the end of inflation is 
\begin{equation}
\begin{aligned}
    n_i^{\rm end}&= \frac{3H^3_{\rm end}M^2_{\rm i}\Sigma_{\rm i}}{2\pi(1-w_\phi)(\gamma m^{\rm end}_\phi)^2}\\
    & = 10^{37}\,\mbox{GeV}^3 \frac{3\,\Sigma_{\rm i}}{2\pi(1-w_\phi)\gamma^2}\left(\frac{H_{\rm end}}{10^{13}\,\mbox{GeV}}\right)^3\\
    &~~~~~~~~~~\times \left(\frac{M_i}{10^{12}\,\mbox{GeV}}\right)^2\left(\frac{10^{13}\,\mbox{GeV}}{m^{\rm end}_\phi}\right)^2 .
    \end{aligned} 
\end{equation}
\begin{figure*}
          \begin{center}      \includegraphics[width=16.0cm,height=11.0cm]{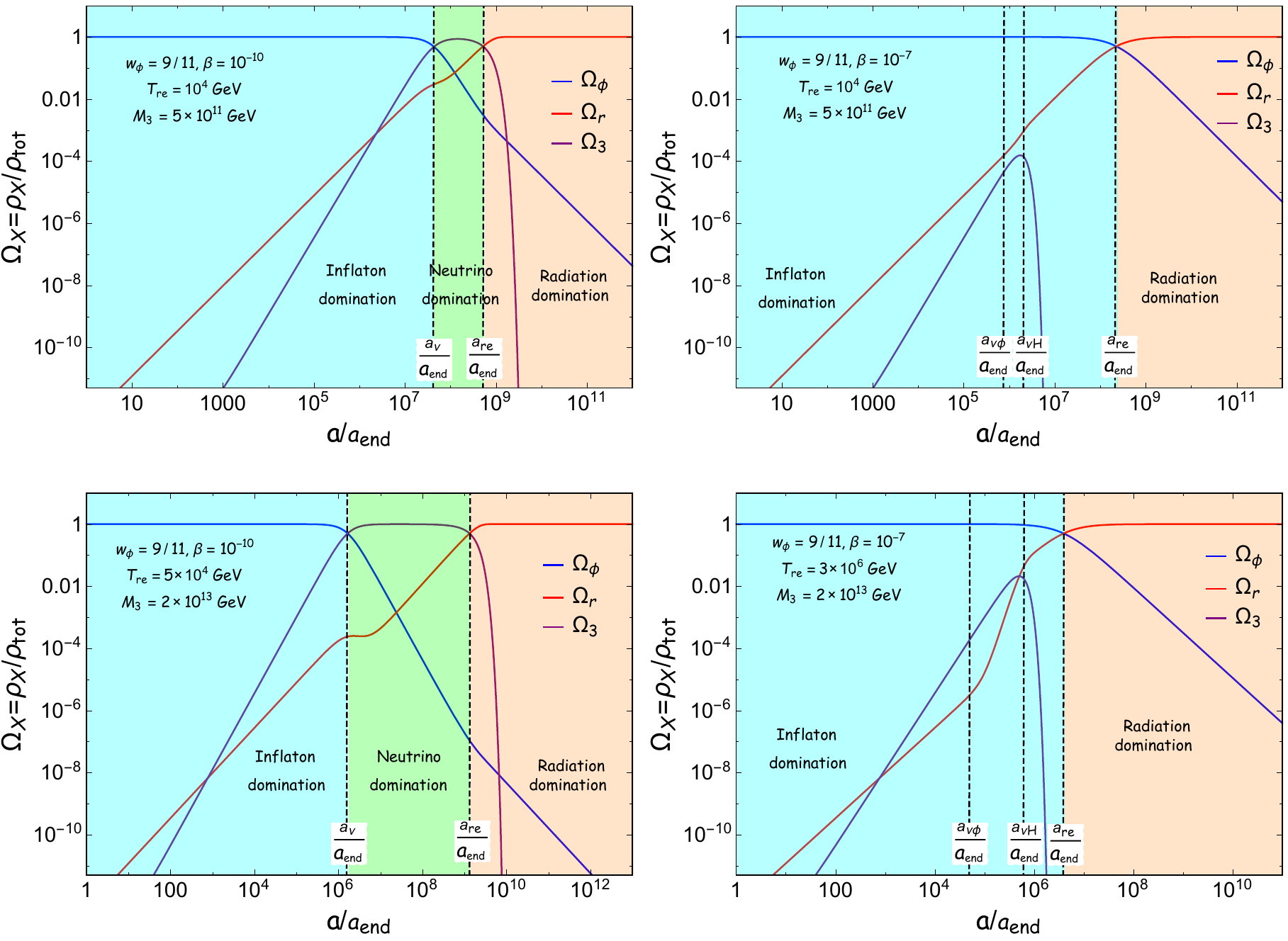}
         \caption{The evolution of the normalized energy densities $\Omega_X=\rho_X/\rho_{\rm tot}$ as a function of the scale factor for the different combinations of $(\beta\,,M_3)$ with $\wphi=9/11$. The left plot corresponds to the neutrino dominating case $(\beta<\beta^{\rm c}_\nu)$ and the right plot corresponds to the neutrino heating case $(\beta>\beta^{\rm c}_\nu)$.}
          \label{density}
          \end{center}
      \end{figure*}
Immediately after the inflation ends, we can observe that approximatly $10^{37}$ number of neutrinos of mass $10^{12}$ GeV are produced per unit volume for a standard high scale inflationary model.  

\subsection{GRe: Estimating critical $\beta^c_\phi$ for neutrino reheating}
To facilitate our discussion forward we begin by stating few basic results of the standard gravitational reheating scenario \cite{Haque:2022kez}. If we put $\Gamma_3=0$ in Eq. \ref{rhor}, then it simply decodes the standard GRe scenario where reheating is purely gravitationally driven with the decay width
 \begin{equation}
\Gamma_{\rm\phi\phi\rightarrow hh}=\frac{\gamma\,m_\phi(t)\,\rho_\phi(t)}{\pi\,(1+w_\phi)M^4_p}\Sigma_{\rm R}\,, 
 \end{equation}
 where $\Sigma_{\rm R}=\sum^{\infty}_{\nu=1}\nu\,\lvert\mathcal P^{2n}_\nu \rvert^2$ and its numerical values given in Table-\ref{fouriersum}. During this, the radiation production from inflton via gravitational scattering can be obtained from Eq.(\ref{rhor}) as,
 \begin{equation}\label{rhorg}
   \rho_{\rm r}(a)\simeq\frac{9\,\gamma \,H^3_{\rm end}\,m^{\rm 
    end}_{\rm\phi}\,\Sigma_{\rm R}}{4\,\pi\,(1+15w_{\rm\phi})}\left(\frac{a}{a_{\rm end}}\right)^{-4}\, .
\end{equation}
In the present scenario where $\Gamma_3\neq0$, radiation can also be produced from neutrino decay. However, if the neutrino decay width is very large, the radiation produced from neutrino decay will never surpass the radiation produced by the inflaton. This is due to the mass suppression of the right handed neutrino production. Therefore, one finds a critical value of the neutrino coupling parameter $\beta$, below which we found neutrino to control the reheating process,
\begin{equation}
\begin{aligned}
    \beta^{c}_\phi\simeq&\left(\frac{9\gamma \Sigma_{R}}{3\pi(1+15w_{\rm\phi}) }\frac{m^{\rm end}_{\rm\phi}}{M_{\rm p}}\right)^{\frac{-3(1+w_\phi)}{4}}\left(\frac{M_{\rm p}}{H_{\rm end}}\right)^{\frac{7+9w_\phi}{4}}\\
  &\quad\times\left(\frac{n_{\rm 3}^{\rm end}}{M^3_{\rm p}}\right)^{\frac{3(1+w_{\rm\phi})}{4}}\left(\frac{M_{\rm p}}{M_3}\right)^{\frac{-(1+3w_{\rm\phi})}{4}}\delta^{3w_\phi-1} ,
    \end{aligned}
\end{equation}
where
 \begin{equation}
     \begin{aligned}
    \delta=\left(\frac{5+3w_\phi}{1+w_\phi}\right)^{\frac{3(1+w_\phi)}{4(1-3w_\phi)}}\left(\frac{1}{12\pi(1+w_\phi)}\right)^{\frac{1}{2(1-3w_\phi)}} .
     \end{aligned}
 \end{equation}
If $\beta \gg \beta^c_\phi$, neutrino, after being produced from inflaton will immediately decay and cannot control the reheating dynamics anymore. In such a case, reheating is predominantly controlled by gravitational scattering, which is the usual gravitational reheating scenario first proposed in  \cite{Haque:2022kez}. Therefore, we will consider the $\beta < \beta^c_\phi$ though out.

As stated earlier, this reheating scenario is nearly ruled out from the $\Delta N_{\rm eff}$ constraints considering extra relativistic degrees of freedom accounted from the PGWs \cite{Haque:2022kez}. Nevertheless, for our later purpose, let us quote the expression of the reheating temperature $T_{re}$ for the GRe scenario 
 \begin{equation}
 \begin{aligned}
    & \tre^{\rm GRe}\simeq0.5\,M_{\rm p}\left(\frac{9\gamma\,\Sigma_{R}}{3\pi(1+15w_{\rm\phi}) }\frac{m^{\rm end}_{\rm\phi}}{M_{\rm p}}\right)^{\frac{3+3w_\phi}{4(3w_{\rm\phi}-1)}}\\
&\quad\quad\quad\times\left(\frac{H_{\rm end}}{M_{\rm p}}\right)^{\frac{9w_\phi+1}{4(3w_\phi-1)}}
     \end{aligned}
     \label{gretre}
 \end{equation}
Taking into account BBN constraints on PGW, GRe turned out to be consistent only for $\wphi \simeq 1$, and the associated reheating temperature is predicted to be $\tre^{\rm GRe} \simeq 10^5$ GeV.

\subsection{Two possible scenarios}
Out of the three heavy right-handed neutrinos, we have assumed the decay width of the third one $\Gamma_3 \ll \Gamma_2,\Gamma_1$. Therefore, due to its longer survival probability, $\nu^3_{\rm R}$ and its subsequent decay into radiation will compete with the inflaton decay. Therefore, during reheating, we will discuss about three major players; inflaton and $\nu^3_{\rm R}$ and radiation. 
Before it decays into the SM particles, $\nu^3_{\rm R}$ will evolve as non-relativistic matter, 
\begin{equation}\label{rhoN3s}
\begin{aligned}
    \rho_{3}(a)&= n_3 M_3\simeq\frac{3H^3_{end}M^3_{3}\Sigma_{3}}{2\pi(1-w_\phi)(\gamma\,m^{\rm end}_\phi)^2} \left(\frac{\aend}{a}\right)^{3}
    \end{aligned}
\end{equation}
Hence, for any value of $w_{\rm\phi}>0$,  inflaton $  \rho_\phi=a^{-3(1+w_\phi)}$, dilutes faster than the $\nuth$, it can dominate at $a=a_{\nu}$ where $\rho_{\rm\phi}\sim\rho_{3}$.
\begin{equation}\label{AphiN}
\begin{aligned}
    &\frac{a_\nu}{\aend} \simeq\left(\frac{2\pi(1-w_{\rm\phi})\gamma^2}{\Sigma_{3}}\right)^{1/3w_\phi}\left(\frac{M_{\rm p}}{H_{\rm end}}\right)^{1/3w_\phi}\\
    &\quad\quad\quad\quad\times\left(\frac{m^{\rm end}_\phi}{M_{\rm p}}\right)^{2/3w_\phi}\left(\frac{M_{\rm p}}{M_3}\right)^{1/w_\phi}.
    \end{aligned}
\end{equation}
 Once $\nu^3_{\rm R}$ dominates over the inflaton, its subsequent decay into SM fields would populate the universe. One, therefore, expects a new critical value of $\beta =\beta^c_\nu$ below which $\nu^3_{\rm R}$ would survive long, and the universe becomes neutrino-dominated over the inflaton at some intermediate period before the conclusion of the reheating. If $\Gamma_{\rm 3} = \beta^2 M_{3}/(8\pi) \sim H(a_{\nu})$, one  obtains following expression for the new critical value as, 
 \begin{equation}
\begin{aligned}
\beta^{\rm c}_\nu &=\left(\frac{8\pi H(a_\nu)}{M_3}\right)^{1/2} \\ 
&\simeq 22.3\left(\frac{H_{\rm end}}{10^{13}\, \mbox{GeV}}\right)^{1/2}\left(\frac{5\times10^{11}\,\mbox{GeV}}{M_3}\right)^{1/2}\\
&\quad\quad\times\left(\frac{a_\nu}{a_{\rm end}}\right)^{-\frac{3(1+w_\phi)}{4}}\,.
\end{aligned}
\end{equation}
Therefore, for any value of $\beta \leq \beta^{\rm c}_\nu$, the $\nu_R^3$ will survive long enough such that the universe becomes neutrino (matter) dominated, say at the scale factor $a=a_\nu$. Then, after its subsequent decay into Standard Model particles will complete the reheating process. For example, we get $\beta^{\rm c}_\nu \simeq 10^{-9}\,(10^{-8})$ for $M_{\rm 3}=5\times 10^{11}\,(2\times10^{13})$ GeV for $w_\phi=9/11$. In the right plot in Fig.(\ref{density}), we indeed see that $\nu^3_{\rm R}$ domination ceases to exist when $\beta = 10^{-7}>\beta^{\rm c}_\nu$. On the other hand, for $\beta = 10^{-10}<\beta^{\rm c}_\nu$, the universe undergoes a neutrino-dominated phase after inflaton domination as depicted in the left panel of the Fig. (\ref{density}).

Depending on $\beta$, interesting possibilities arise for $ \beta^c_\phi\geq\beta\geq\beta^c_\nu$. For this condition, inflaton still dominates throughout the entire reheating period. However, neutrinos can populate the thermal bath more efficiently than the inflaton. 

In summary, if the coupling is very large $\beta\gg\beta^c_\phi$, neutrinos produced from inflaton will decay immediately, leaving no traces during the entire reheating process. This is what a GRe scenario is all about and has been studied before \cite{Haque:2022kez}.

In this paper, we advocate two remaining possibilities depending on the $\nu^3_{\rm R}$ decay width: (1) {\bf Neutrino dominating}: Heavy neutrinos can dominate the universe as a matter-like component over inflaton at some point during reheating and then decay to complete the reheating process, (2) {\bf Neutrino heating}: Inflaton is the dominant energy component during the entire reheating process; however, the neutrino decay controls the reheating temperature.

\subsubsection{ \bf{Neutrino dominating: $\mathbf\beta \leq \beta^{\rm c}_\nu$ }}
After the end of inflation, the gravitationally produced neutrino $\nu^3_{\rm R}$ will start to populate the universe. It is the long-lived  that dominates over the inflaton after $a=a_{\rm\nu}$, and the universe turned into matter-dominated with $\rho_3 \propto a^{-3}$ (see Eq.\ref{rhoN3s}), then the Hubble parameter evolves as 
 \begin{equation}
     H(a)=H_{\rm end}\left(\frac{a_{\rm\nu}}{a_{\rm end}}\right)^{-\frac 3 2(1+w_{\rm\phi})} \left(\frac{a}{a_\nu}\right)^{-\frac{3}{2}}~~~~~~a \geq a_{\rm\nu}\,.
 \end{equation}
During this period, radiation will naturally be populated by the $\nu_{\rm R}^3$ decay, and reheating will be completed at $\Gamma_3\sim H(\are)$. Considering neutrino dominated universe and utilizing   Eq.(\ref{rhor}), radiation evolves as  
  \begin{equation}{\label{xyz}}
    \rho_{\rm r}(a \geq a_{\rm\nu})\simeq\frac{\beta^2M^2_{3}\,n_{\rm 3}(a_{\rm\nu})}{20\,\pi\,H(a_{\rm\nu})} \left(\frac{a}{a_{\rm\nu}}\right)^{-3/2} ,
\end{equation}
which is independent of $w_{\rm\phi}$ as expected. However, it is important point to remember that during pre-neutrino domination $(a<a_{\rm\nu})$, the radiation was controlled by inflaton via its gravitational decay, and hence, the maximum radiation temperature lies at the beginning of reheating and determined by inflaton. In our subsequent discussion we will see this will play an interesting role in the context of dark matter production during reheating. The maximum radiation temperature will be,
\begin{equation}
    T^{\max}_{\rm rad}\simeq(10^{11}\,\mbox{GeV})\left(\frac{H_{\rm end}}{10^{13}\,\mbox{GeV}}\right)^{3/4}\left(\frac{m^{\rm end}_\phi}{10^{13}\,\mbox{GeV}}\right)^{1/4},
\end{equation}
which is of the order of  $\sim 10^{11}-10^{12}$ GeV. 
However, the final reheating temperature is determined by $\nu^3_{\rm R}$ decay, as indeed can be seen in the left plot in Fig. (\ref{density}). Since $\rho_{\phi} \sim a^{-3(1+w_\phi)}$, for successful reheating, one must have $w_\phi\geq1/3$; otherwise, due to slow dilution of inflaton compared to radiation, the universe will become inflaton dominated at a later stage, which we want to avoid. As stated earlier, the reheating ends at $a=a_{\rm re}$ where $\Gamma_{\rm 3} = (3/2) H(a_{\rm re})$ leading to,
\begin{equation}{\label{areN}}
    \frac {a_{\rm re}}{a_{\rm\nu}}=\left(\frac{\beta^2}{12\pi}\frac{M_{\rm 3}}{H(a_{\rm\nu})}\right)^{-2/3} .
\end{equation}
In equilibrium the radiation temperature $T$ relates to its energy density as $\rho_{\rm r}=\epsilon\, T^4$, with $\epsilon = \pi^2 g_{*\rm r}/30$. Where $g_{*\rm r}$ is the relativistic degrees of freedom associated with the thermal bath. Utilizing this
the radiation temperature evolves during $a_{\rm\nu}\leq a\leq\are$, as [cf. Eq.\ref{xyz}]
\begin{equation}{\label{n38}}
    T(a)\simeq \tre\left(\frac{a}{\are}\right)^{-3/8}
\end{equation}
where the reheating temperature $\tre$,
\begin{equation}{\label{TreN}}
\begin{aligned}
    T_{\rm re}&=\beta\left(\frac{2}{3\,\epsilon}\right)^{1/4}\left(\frac{M_{\rm 3}\,M_{\rm p}}{{8\,\pi}}\right)^{1/2}\,,\\
    &\simeq  10^{14}\beta\left(\frac{M_3}{5\times 10^{11}\mbox{GeV}}\right)^{1/2}\mbox{GeV} .
    \end{aligned}
\end{equation}
We observed that reheating temperature behaves as $\tre \propto \beta\,M_3^{1/2}$.
For the lowest possible reheating temperature $\tre = T_{\rm  BBN} = 4$ MeV, one gets $\beta \simeq 4\times10^{-17}$ for $  M_3=5\times10^{11}$ GeV. 
On the other hand for the neutrino dominating case, the maximum reheating temperature could be achieved at $\beta=\beta^{\rm c}_\nu$ for a given $M_3$ as
\begin{equation} \label{trebc}
    \tre(\beta^{\rm c}_\nu) = 10^{14}\beta^{\rm c}_\nu \left(\frac{M_3}{5\times 10^{11}\mbox{GeV}}\right)^{1/2}\mbox{GeV}\,.
\end{equation}
For example given $w_\phi = 9/11\,(1/2)$ and $M_3=5\times10^{11}$ GeV, one obtains the the maximum reheating temperature as $10^{5}\,(34)$ GeV.  

In summary, for the neutrino-dominated case, before it decays, the universe will go through inflaton domination (ID) $\to$ RHN domination $\to $ radiation domination (RD) era (see left plot in Fig.\ref{density}).  The evolution of the radiation energy density in those phases can be written as  

\begin{equation}
  \rho_{\rm r} \propto \left\{ \begin{array}{lll}  
   a^{-4} & {\rm for }~\aend <a<a_\nu & :{\rm ID} \\
   a^{-3/2} & {\rm for }~  a_\nu <a<\are & :{\rm RHN~domination} \\
   a^{-4} & {\rm for }~ a > \are  & :{\rm RD} .\\
   \end{array} \right.
\end{equation}

\subsubsection{\bf{Neutrino heating :} $\mathbf \beta^{c}_\nu\leq\beta\leq\beta^{c}_\phi$}
This is the most interesting case we have encountered. Despite neutrinos being a subdominant component, the decay of RHNs is the primary source of the thermal bath than the inflation. Neutrinos to become the dominating source of radiation the condition $M_3\Gamma_{3}\,n_3>\Gamma_{\rm\phi\phi\rightarrow hh}(1+w_{\rm\phi})\rho_{\rm\phi}$ must be satisfied, and that happens at $a=a_{\rm\nu\phi}$ satisfying 
\begin{equation}
\left(\frac{a_{\rm\nu\phi}}{\aend}\right)^{3+9\wphi}\simeq\frac{48\,(1-\wphi)\,\pi\,\gamma^3}{\beta^2}\frac{\Sigma_{\rm R}}{\Sigma_3}\frac{\hend}{M_3}\left(\frac{m^{\rm end}_\phi}{M_3}\right)^3
\end{equation}
Therefore, after $a>a_{\nu\phi}$ ignoring the inflaton contribution, the radiation components solely coming from the neutrino ($\nu_{\rm R}^3$) decay and it's computed as (solving Eq.(\ref{rhor})
\begin{equation}\label{rhorphi}
    \rho_{\rm r}(a>a_{\nu\phi})\simeq\frac{2\beta^2_{\rm 3}M^2_{\rm 3}n^{\rm end}_{\rm3}}{8\pi(5+3w_{\rm\phi})H_{end}}\left(\frac{a}{a_{\rm end}}\right)^{-\frac{3(1-w_{\rm\phi})}{2}} .
\end{equation}
\iffalse
\begin{figure*}
          \begin{center}
\includegraphics[width=15.50cm,height=5.50cm]{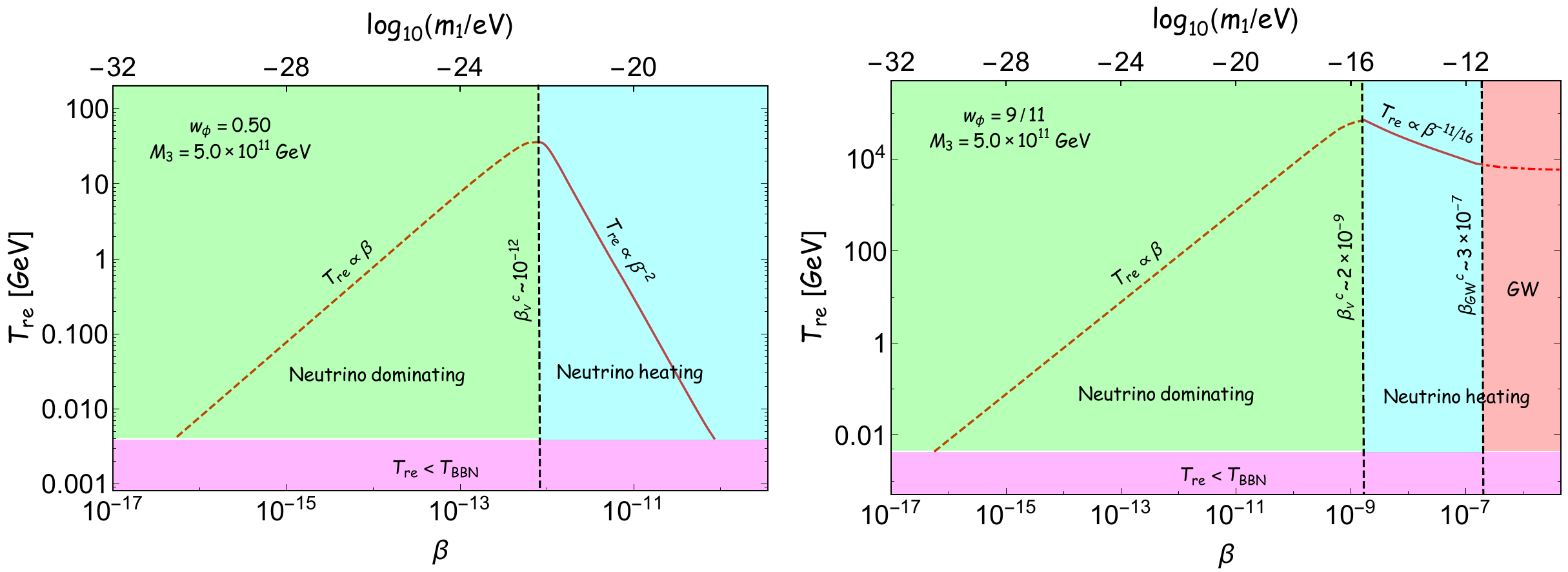}
         \caption{ Variation of reheating temperature $T_{\rm re}$ as a function of $\beta$ for two different $w_{\rm\phi}=1/2$ (right plot) and $w_{\rm\phi}=9/11$ (left plot) for $M_{\rm3}=5\times 10^{11}$ GeV. BBN bounds exclude the magenta-shaded regions, and red-shaded regions (left plot) are excluded by an excess of gravitational wave.}
          \label{Tre}
          \end{center}
      \end{figure*}
      \else
      \begin{figure*}
          \begin{center}
\includegraphics[width=17.0cm,height=10.5cm]{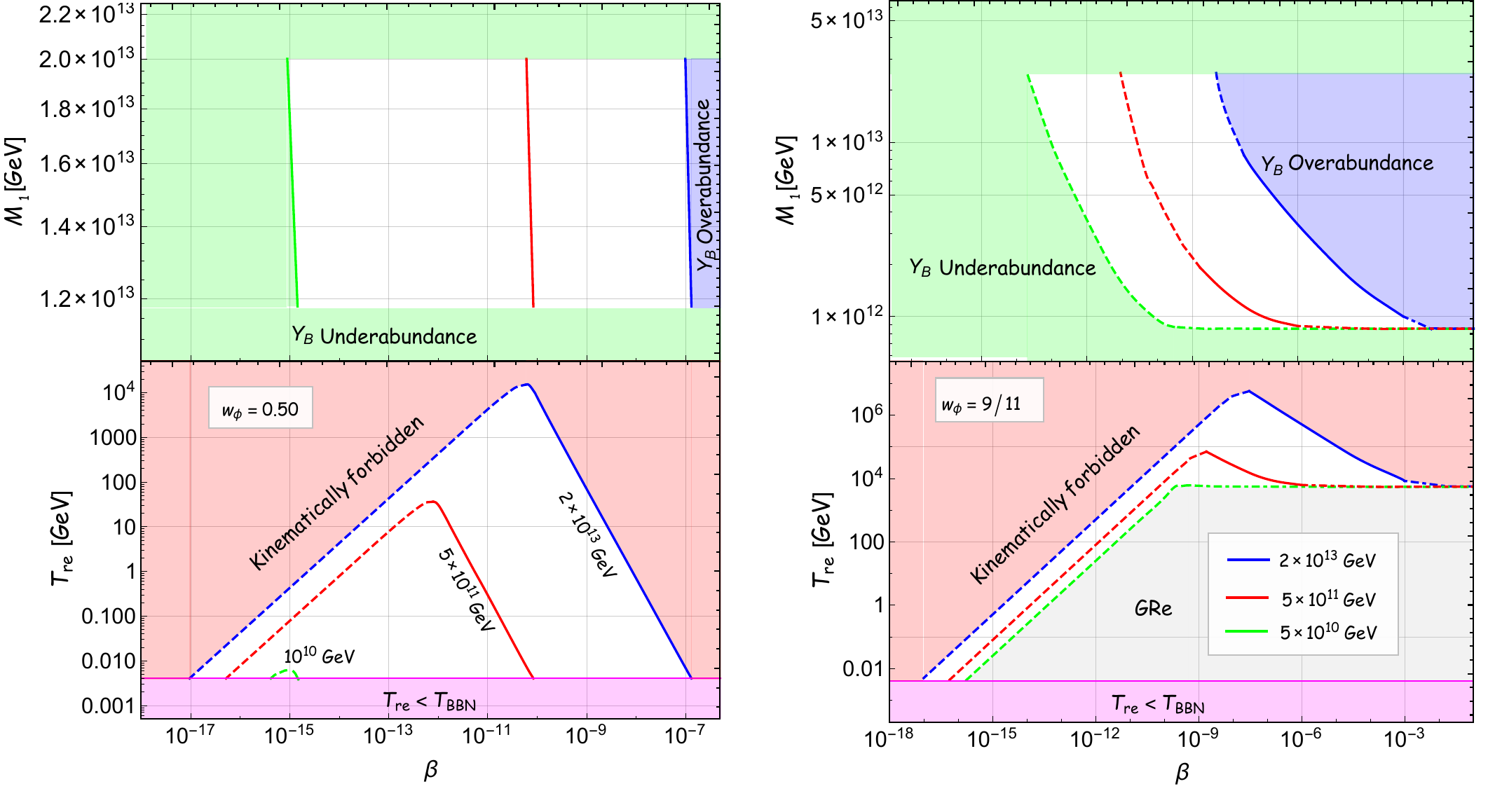}\quad
 \caption{\textit{\bf lower panel} :  $T_{\rm re}$ as a function of $\beta$ for different  $\nu^3_{\rm R}$ mass $M_{\rm 3}$. The dashed line corresponds to the neutrino-dominating case, and the solid line corresponds to the neutrino-heating case. The dot-dashed lines are excluded by an excess production of gravitational waves from $\Delta N_{\rm eff}$ constraints. The magenta-shaded region corresponds to $\tre < T_{\rm BBN}$. The red-shaded region is kinematically forbidden. \textit{\bf upper panel} : we have plotted the contour of the observed baryon asymmetry $Y_{\rm B}\simeq 8.7\times10^{-11}$ in $ \beta$-$M_1$ plane. The different color line corresponds to different $M_3$, the same as the bottom plot. The white regions are the allowed regions where the observed baryon asymmetry is possible, the blue-shaded regions are the overabundance regions, and the green-shaded regions correspond to the underabundance regions.}
          \label{pre}
          \end{center}
      \end{figure*}
This evolution will continue till the instant, say at $a=a_{\rm \nu H} \leq \are$, for which $\Gamma_{\rm 3}\sim H$ is satisfied. We have
 \begin{equation}
     \begin{aligned}
         a_{\rm \nu H}/a_{\rm end}=\left(\frac{\beta^2}{12\pi(1+w_{\rm\phi})}\frac{M_{\rm 3}}{H_{\rm end}}\right)^{\frac{-2}{3(1+w_{\rm\phi})}} .
     \end{aligned}
 \end{equation}
Therefore, within $a_{\nu\phi}<a< a_{\rm\nu H}$ the radiation temperature will evolve as
\begin{equation}\label{rhorphi}
    T \simeq \left(\frac{15\,\beta^2_{\rm 3}M^2_{\rm 3}n^{\rm end}_{\rm3}}{2\,g_{\star\rm r}\, \pi^2 (5+3w_{\rm\phi})H_{end}} \right)^{\frac14} \left(\frac{a}{a_{\rm end}}\right)^{-\frac{3(1-w_{\rm\phi})}{8}}
\end{equation}
Such equation-of-state (EoS) dependent temperature evolution will continue till the neutrinos decay completely into radiation. After this instant $a_{\rm \nu H}(<a_{\rm re})$, there will be no additional entropy injected into the thermal bath, and hence the radiation component simply falls as $a^{-4}$ with $T\propto a^{-1}$. Finally the reheating temperature will set by the condition at $\rho_\phi(a_{\rm re})\sim \rho_{\rm r}(a_{\rm re})$ (see right plot in Fig. (\ref{density})). However, it is important to note that such possible arises only for the inflaton EoS $w_\phi>1/3$. The endpoints of reheating $\are$ and the reheating temperature $\tre$ are expressed as, 
 \begin{equation}{\label{Trephi}}
\begin{aligned}
  & \frac{a_{\rm re}}{a_{\rm end}}=\left(\frac{8\pi(5+3w_\phi)}{\beta^2}\frac{H^3_{\rm end}M^2_p}{n^{\rm end}_3M^2_3}\right)^{1/3w_\phi-1}\left(\frac{a_{\rm\nu H}}{a_{\rm end}}\right)^{\frac{-(5+3w_{\rm\phi})}{2(3w_{\rm\phi}-1)}}\,,\\
    &T_{\rm re}\simeq 0.5M_{\rm p}\,\beta^{\frac{-1}{3w_\phi-1}}\left(\frac{M_{\rm p}}{H_{\rm end}}\right)^{\frac{3}{2(3w_\phi-1)}}
    \left(\frac{n^{\rm end}_3}{M^3_{\rm p}}\right)^{\frac{3(1+w_\phi)}{4(3w_\phi-1)}}\\
    &\quad\quad\quad\times\left(\frac{M_{\rm p}}{M_{\rm 3}}\right)^{\frac{1+3w_\phi}{4(1-3w_\phi)}}\delta\,.
     \end{aligned}
 \end{equation}
Reheating temperature depends nontrivially on the neutrino coupling and the mass of the $\nu^3_{\rm R}$, $\tre\propto\beta^{\frac{-1}{3w_{\rm\phi}-1}}\, M_{\rm 3}^{\frac{7+9w_\phi}{4(3w_\phi-1)}}$. As an example, once we fixed $M_{\rm 3}$, for $w_\phi=1/2$, the $\tre\propto\beta^{-2}$, and for $w_\phi=9/11$, we got $\tre\propto \beta^{-11/16}$ which we also confirm numerically and represented in the bottom panel of the Fig.\ref{pre} by solid lines. For this neutrino heating case, the maximum $\tre$ is associated with $\beta = \beta^{\rm c}_\nu$, and it is the same as the neutrino dominating case. The minimum value of $\tre$ is set by either BBN energy scale or purely gravitational reheating case, which is only applicable for $\wphi>0.65$.

In summary, for the neutrino heating case, the universe will go through ID $\to$ RD.  However, as discussed during the inflaton domination, radiation is initially sourced by the inflaton decay and then controlled by the decay of RHNs, leading to the following evolution history of the thermal bath (see the right plot in Fig.\ref{density}),
\begin{equation}
  \rho_{\rm r} \propto \left\{ \begin{array}{lll}  
   a^{-4} & {\rm for }~\aend <a<a_{\nu\phi} & :{\rm ID} \\
   a^{-\frac{3(1-w_{\rm\phi})}{2}} & {\rm for }~  a_{\nu\phi} < a < a_{\rm \nu H} & :{\rm ID} \\
   a^{-4} & {\rm for }~ a_{\rm \nu H} < a < \are  & :{\rm ID} \\
   a^{-4} & {\rm for }~ a > \are  & :{\rm RD} .\\
   \end{array}\right.
\end{equation}

Considering details of the different reheating histories discussed above, we now depict the parameter regions of $\{\beta\,, M_3\,, w_\phi\,, \tre\}$ where different histories are realized. In the bottom panel of Fig.\ref{pre}, we have plotted $\tre$ as a function of $\beta$ within the allowed range of $M_3$. The different color lines correspond to different values of $M_3$. The corresponding mass range is taken within ($10^{10}, 2\times10^{13}$) GeV for $\wphi=1/2$ and, ($5\times10^{10},2\times10^{13}$) GeV for $\wphi=9/11$. From Fig.\ref{pre}, for a given $\beta$, $T_{\rm re}$ increases with increasing  mass of $\nu^3_{\rm R}$ $M_{3}$, due to its increase of entropy injection. However, it worth remembering that the gravitational production rate of neutrino is $\propto \left(1-\frac{4M^2_{3}}{k^2\gamma^2m^2_{\rm\phi}}\right)^{3/2}$. As a result, for $M_3 \approx \gamma\, m^{\rm end}_{\rm\phi}\simeq2\times 10^{13}$ GeV kinetic suppression leads to negligible neutrino production, and consequently reheating would not be possible due to significant kinematic suppression (depicted by red shaded region). 
On the other hand, the lowest possible mass $M_3$ can be determined from the condition $\tre(\beta^{\rm c}_{\rm\nu})\sim\tre^{\rm min}$,
and the final expression of $M^{\rm min}_3$ is
\begin{equation}
\begin{aligned}
    M^{\rm min}_3\simeq&\mp\left(\frac{2\pi\,(1-\wphi)\gamma^2}{\sum_3}\right)^{1/3}\left(\frac{m^{\rm end}_\phi}{\mp}\right)^{2/3}\\
   &\,\,\, \left(\frac{\mp}{H_{\rm end}}\right)^{\frac{3\wphi+1}{3(1+\wphi)}}\left(\frac{\tre^{\rm min}}{\mp}\right)^{\frac{4\wphi}{3(1+\wphi)}}\,.
    \end{aligned}
\end{equation}
Using $\wphi=0.50\,(9/11)$, one can find $M^{\rm min}_3\simeq1\,(5)\times10^{10}\,$GeV.

 The primary gravitational waves (PGWs) are the evolution of the tensor perturbation without any source term, and they can act as a good probe of the early universe. Once the tensor perturbations are generated due to vacuum fluctuation during inflation, they pass through different phases until we observe it today and put imprints of the background evolution on the spectrum. In the following section, we show how constraints from extra relativistic degrees of freedom at the point of BBN considering PGWs put restrictions on our proposed model of reheating.

\section{Constraints from $\Delta N_{\rm eff}$ considering PGWs }{\label{sc3}}
PGWs is one of the profound predictions of inflationary paradigm \cite{Grishchuk:1974ny,Guzzetti:2016mkm,Starobinsky:1979ty}. Due to extremely weak coupling, it can pass through the universe almost unhindered and hence acts as a unique probe of the very early universe such as reheating phase \cite{Mishra:2021wkm,Benetti:2021uea,Haque:2021dha,Maity:2024cpq,Vagnozzi:2023lwo,Vagnozzi:2020gtf}. The PGW amplitude and its spectral evolution are expected to be sensitive to the energy scale of the inflation and the post-inflationary phases. Particularly during the reheating phase, it has been observed that the modes between $k_{\rm re} < k < k_{\rm end}$ which re-enter the horizon during reheating, PGW spectrum shows a tilt depending on the background equation of state.  Once we fixed the inflationary energy scale and $w_\phi$, $T_{\rm re}$ regulates the duration of the tilted spectrum. The tilted spectrum's duration extends as we reduce $T_{\rm re}$. Nonetheless, if one considers the extra relativistic degrees of freedom arising from the GWs, the lesser value of $T_{\rm re}$ for a stiff equation can be constrained. We are going to discuss how small $T_{\rm re}$, after we fix the EoS and the inflation's energy scale, we can consider being consistent with the $\Delta N_{\rm eff}$. The additional relativistic degrees of freedom at the moment of BBN or CMB decoupling are denoted by $\Delta N_{\rm eff}$. The expression for $\Delta N_{\rm{eff}}$ in the case of the GWs takes the following form \cite{Jinno:2012xb}
\begin{equation}
    \Delta N_{\rm{eff}}=\frac{\rho_{_{\rm GW}}}{\,\rho_{\nu}}
   =\frac{8}{7}\left(\frac{11}{4}\right)^{\frac{4}{3}}\frac{\rho_{_{\rm GW}}}{\rho_{\rm \gamma}}\,,
   \label{Eq: neff}
\end{equation}
where $\rho_{\rm \gamma}$ and $\rho_\nu$ represent the photon energy density and energy density of single SM neutrino species, accordingly. One can use the relation between the neutrino and photon temperature, $T_{\rm \nu} = \left({4}/{11}\right)^{{1}/{3}}\, T_{\rm \gamma}$ and find the restriction on the current GW energy density as
\begin{align} 
  \int_{k_{\rm re}}^{k_{\rm end}}\frac{dk}{k}\Omega_{\rm GW}h^2(k)\leq \frac{7}{8}\left(\frac{4}{11}\right)^{4/3}\Omega_{\rm \gamma}h^2\,\Delta{\rm N_{\rm eff}},
  \label{eq: deltaneff}
\end{align}
where $\Omega_{\rm \gamma}h^2\simeq 2.47\times10^{-5}$ is the present day photon relic density. $k_{\rm re}$ and $k_{\rm end}$ represent the wave number that reenters at the end of reheating and inflation, respectively.

We note that in the neutrino dominating case $\beta \leq \beta^{\rm c}_\nu$, due to neutrino domination same as matter, the PGWs follow the well-known red tilted spectral behavior $\propto k^{-2}$ particularly for scales within $\kre \leq k \leq k_\nu$. $k_\nu =a_\nu\,H(a_\nu)$ which enter during reheating at the commencement of neutrino domination,  and $k_{\rm re}=a_{\rm re}H(a_{\rm re})$ which enters the horizon at the end of reheating. On the other hand, higher frequency modes within $\kend \geq k \geq k_\nu $ follow the blue tilted spectrum as  \cite{Haque:2023zhb}, 
\be \label{GW3}
\Omega^{\rm k}_{\rm GW}h^2 \simeq \Omega^{\rm inf}_{\rm GW}h^2 \frac{\Gamma^2 (\frac 5 2)}{\pi}\left(\frac{k_{\rm re}}{k_\nu}\right)^{2} \frac{\mu (w_\phi)}{\pi}\left(\frac{k}{ k_{\rm \nu}}\right)^{-\frac{(2-6\,w_\phi)}{(1+3\,w_\phi)}},
\ee
where, $\mu(w_\phi)=(1+3\omega_\phi)^{\frac{4}{1+3\,w_\phi}}\,\Gamma^2\left(\frac{5+ 3\,w_{\phi}}{2+6\,w_{\phi}} \right)$, which typically assumes ${\cal O}(1)$ value. $k_{\rm end}$ is the largest frequency mode that reenters the horizon at the end of the inflation, which typically contributes to the $ \Delta N_{\rm{eff}}$. Note that if there is no intermediate neutrino (matter) domination such as the case of usual GRe, the form of the above spectrum Eq.\ref{GW3} would be the same with $k_\nu$ replaced by $\kre$. This essentially suggests that due to late matter domination, spectrum in Eq.\ref{GW3} acquires a large suppression factor $({\kre}/{k_\nu})^{\frac{12\,w_\phi}{(1+3\,w_\phi)}} << 1$ for high frequency modes. For example taking $\wphi=9/11$, we obtain $\tre^{\rm GRe}\sim4\times10^{3}$ GeV and $\kre\sim10^{-3}$ Hz.  Whereas taking $M_3=5\times10^{13}$ GeV and keeping the same $\wphi\,,\tre$ as above, $\nu$GRe yields  $k_{\rm\nu}\sim10^{-2}$ Hz.  Utilizing those values, we obtain the suppression factor to be $\sim 10^{-3}$. Therefore, the neutrino dominating case always abides by the $\Delta N_{\rm eff}$ constraint for all equation state $\wphi > 1/3$.

However, the situation gets strikingly different for the neutrino heating case ($\beta^{\rm c}_\nu\leq\beta\leq\beta^{\rm c}_\phi$).
For this neutrino decay controls the reheating temperature, and inflation controls the background during the entire reheating period. For such cases, all the modes that enter the horizon during reheating, the PGWs spectrum we have \cite{Haque:2021dha,Chakraborty:2023ocr}
\be \label{GW1}
\Omega^{\rm k}_{\rm GW}h^2\simeq \Omega^{\rm inf}_{\rm GW}h^2 \frac{\mu (w_\phi)}{\pi}\left(\frac{k}{ k_{\rm re}}\right)^{-\frac{(2-6\,w_\phi)}{(1+3\,w_\phi)}} .
\ee 
The scale-invariant part $\Omega^{\rm inf}_{\rm GW}h^2$,
is fixed  by the inflationary energy scale as follows, 
\be \label{omegainf}
\Omega^{\rm inf}_{\rm GW}h^2
= \frac{\Omega_{\rm R} h^2 H_{\rm end}^2}{12 \pi^2 M_{\rm p}^2}\, = 6 \times 10^{-18} \left(\frac{H_{\rm end}}{10^{13}\,\mbox{GeV}}\right)^2.
\ee
Where we used the present radiation abundance $\Omega_{\rm R} h^2= 4.16\times10^{-5}$. One can observe that the spectrum is blue tilted for $w>1/3$, and therefore, its magnitude would be maximum for the mode $k=k_{\rm end}$. This is where the BBN bound related to the effective number of relativistic degrees of freedom comes into play, and this translates into the bound on PGW spectrum as $\Omega^{\rm k_{end}}_{\rm GW}h^2\leq1.7\times10^{-6}$ \cite{Pagano:2015hma,Yeh:2022heq,Planck:2018vyg} considering $\Delta N_{\rm eff}=0.28$ \cite{Planck:2018vyg}.
Utilizing the Eqs. \ref{eq: deltaneff} and \ref{GW1}, we obtain the following contraint,
\begin{equation}{\label{gwbbn1}}
    \left(\frac{k_{\rm end}}{k_{\rm re}}\right)^{-\frac{2(1-3w_\phi)}{1+3w_\phi}}\leq\frac{4.85\pi}{\mu(w_\phi)}\left(\frac{M_{\rm p}}{H_{\rm end}}\right)^2\,.
\end{equation}
The relation between the two scales $(k_{\rm end},\,k_{\rm re})$ can be further expressed in terms of reheating temperature $\tre$,
as follows
\begin{equation}{\label{kendkre1}}
\frac{k_{\rm end}}{k_{\rm re}}=\left(\frac{H_{\rm end}}{3.3\, M_{\rm p}}\right)^{\frac{1+3w_\phi}{3(1+w_\phi)}}\left(\frac{M_{\rm p}}{T_{\rm re}}\right)^{\frac{2(1+3w_\phi)}{3(1+w_\phi)}}\,,
\end{equation}
 where we have used the relation between $a_{\rm re}$ and $T_{\rm re}$ 
\begin{equation}{\label{are1}}
    \frac{a_{\rm re}}{a_{\rm end}}=\left(\frac{H_{\rm end}}{3.3\,M_{\rm p}}\right)^{\frac{2}{3+3w_{\rm\phi}}}\left(\frac{M_{\rm p}}{T_{\rm re}}\right)^{\frac{4}{3+3w_{\rm\phi}}} ,
\end{equation}
The expressions of $\kend$ in terms of $\tre$ can be written as,
\begin{equation}
   \begin{aligned}
       k_{\rm end}&=3.3\,\mp\left(\frac{43}{11\,g_{\star \rm r}}\right)^{1/3}\left(\frac{T_0}{\tre}\right)\\
       &\,\quad\quad\times\left(\frac{H_{\rm end}}{3.3\,\mp}\right)^{\frac{1+3w_\phi}{3+3w_\phi}}\left(\frac{\tre}{\mp}\right)^{\frac{4}{3+3w_\phi}}\,,\\
       &\simeq(3\times10^{\frac{3(8+w_\phi)}{1+w_\phi}}\,\mbox{Hz})\times\left(\frac{H_{\rm end}}{10^{13}\,\mbox{GeV}}\right)^{\frac{1+3w_\phi}{3+3w_\phi}}\\
       &\quad\quad\times\left(\frac{\tre}{\mbox{GeV}}\right)^{\frac{1-3w_\phi}{3+3w_\phi}}\,.
   \end{aligned} 
\end{equation}
Using the last form of above equation and Eq.\ref{kendkre1}, one can find 
\begin{eqnarray}
\kre\simeq 1.2\times10^{-9}\,\left(\frac{\tre}{4 ~\mbox{MeV}}\right)\, \mbox{Hz}.
\end{eqnarray}
Finally, upon substitution of Eq.\ref{kendkre1} into Eq.\ref{gwbbn1}, we obtain a
lower bound on the reheating temperature defined, particularly when
the inflaton equation of state $w_\phi>1/3$ as,
\begin{equation}{\label{tregw1}}
    \tre>0.5M_{\rm p}\left(\frac{4.85\pi}{\mu(w_\phi)}\right)^{\frac{3(1+w_\phi)}{4(1-3w_\phi)}}\left(\frac{H_{\rm end}}{M_{\rm p}}\right)^{\frac{1+3w_\phi}{3w_\phi-1}}=T^{\rm GW}_{\rm re} .
\end{equation}
We symbolize this
new lower limit on reheating temperature from PGW as $\tre^{\rm GW}$. Setting the above temperature with the BBN energy scale
$\tre^{\rm GW}\sim4\,\mbox{MeV}$, we can see that the BBN bound of
PGWs only important when $w_\phi\geq0.60$, considering inflation energy scale $H_{\rm end}\sim 10^{13}$ GeV. To this end, let us remind ourselves that such bound also exists for purely gravitational
reheating cases, which satisfies the BBN bound only for $w_\phi \simeq 1$ \cite{Haque:2022kez,Barman:2022qgt}. However, for the present case, due to an additional source of radiation, we achieve a much higher reheating temperature compared to that of the GRe. As we already mentioned, increasing reheating temperature for a fixed $w_\phi$ and $H_{\rm end}$ reduces the duration of the tilt and makes our presented scenario consistent in the domain $1/3<w_\phi\leq1$.  

We further point out that this lower bound on the reheating temperature naturally sets a critical value of the coupling parameter $\beta$ by equating Eqs. \ref{tregw1} and \ref{Trephi},
\begin{equation}
\begin{aligned}
   & \beta^{\rm c}_{\rm GW}\simeq\delta^{3w_\phi-1}\left(\frac{4.85\pi}{\mu(w_\phi)}\right)^{\frac{3(1+w_\phi)}{4}}\left(\frac{n^{\rm end}_3}{M^3_{\rm p}}\right)^{\frac{3(1+w_\phi)}{4}}\\
&\quad\quad\quad\times\left(\frac{M_{\rm p}}{H_{\rm end}}\right)^{\frac{5+6w_\phi}{2}}\left(\frac{M_3}{M_{\rm p}}\right)^{\frac{1+3w_\phi}{4}}, 
    \end{aligned}
\end{equation}
which turned out to be 
$\beta^{\rm c}_{\rm GW} <\beta^c_{\phi}$. 
As an examples for $w_\phi=9/11$ ans $M_3=5\times10^{11}$ GeV, we numericlly obtained $\beta^{\rm c}_{\rm GW}\simeq10^{-7} < \beta^{\rm c}_\phi\simeq10^{-6}$. Further, the reheating temperature is $\tre^{\rm GW}\simeq8\times10^3$ GeV, which is an order of magnitude larger than the prediction for the usual gravitational reheating case $\tre^{\rm GRe} \sim 10^3\,\mbox{GeV}$. Since, $\beta^{\rm c}_{\rm GW} 
< \beta^c_{\phi}$, therefore, the upper bound of $\beta$ is set by  $ \beta^{\rm c}_{\rm GW}$ when $w_\phi>0.60$, whereas for $w_\phi\leq0.60$, the upper bound on $\beta$ is set by $T_{\rm BBN}\sim4\,\mbox{MeV}$.

From our discussion so far, let us reiterate again that in order to have successful reheating, $\beta$ should be non-vanishing. As stated earlier, this predicts non-zero 
lowest active neutrino mass ($m_1=(\beta\,v)^2/M_3$). In Table-\ref{w05}, \ref{w082}, we have provided the allowed range of $\beta$ along with its predicted lowest active neutrino mass.
\begin{table}[h!]
\scriptsize{
\caption{ Prediction of $\beta$ and the lightest neutrino mass $m_{1}$ (in eV) successive reheating for $w_\phi=0.5$}\label{w05}
\centering
 \begin{tabular}{||c| c| c| c||} 
 \hline
 $M_3 (\mbox{GeV})$ & $ \tre(\beta^{\rm c}_\nu)(\mbox{GeV})$&$\beta_{\rm min} (m^{\rm min}_1)$ & $\beta_{\rm max} (m^{\rm max}_1))$\\ [0.5ex] 
 \hline\hline
 $2\times10^{13}$ &$2\times10^4$ &$1\times10^{-17}$($10^{-34}$)&$1\times10^{-7}$($10^{-14}$) \\
 $5\times10^{11}$& $34.0$ & $5\times10^{-17}$ ($10^{-31}$)&$8\times10^{-11}$ ($4\times10^{-19}$) \\
$10^{10}$& $0.006$ & $4\times10^{-16}$ ($5\times10^{-28}$)&$1\times10^{-15}$ ($7\times10^{-27}$)\\[1ex] 
 \hline
 \end{tabular}}
\end{table}
\begin{table}[h!]
\scriptsize{
\caption{ Prediction of $\beta$ and the lightest neutrino mass $m_{1}$ (in eV) successive reheating  for $w_\phi=0.82$}\label{w082}
\centering
 \begin{tabular}{||c| c| c| c||} 
 \hline
 $M_3 (\mbox{GeV})$ & $ \tre(\beta^{\rm c}_\nu)(\mbox{GeV})$&$\beta_{\rm min} (m^{\rm min}_1)$ & $\beta_{\rm max} (m^{\rm max}_1)$\\ [0.6ex] 
 \hline\hline
 $2\times10^{13}$ &$6\times10^6$ &$6\times10^{-18}$ ($3\times10^{-35}$)&$9\times10^{-4}$($8\times10^{-7}$) \\
 $5.0\times10^{11}$& $7\times10^4$ & $5\times10^{-17}$ ($2\times10^{-31}$)&$1\times10^{-7}$ ($6\times10^{-13}$) \\
$5.0\times10^{10}$& $5\times10^3$ & $2\times10^{-16}$ ($2\times10^{-29}$)&$5\times10^{-10}$ ($2\times10^{-16}$)\\[1ex] 
 \hline
 \end{tabular}}
\end{table}
In the next section, we will discuss where our proposed model lies in the light of recent BICEP/ $Keck$ data together with Planck.
\section{Impact on inflationary parameters $(N_{\rm I},n_{\rm s},\,r)$}{\label{sc4}}
 Along with the reheating phase, the post-reheating history is also important in order to put constrain on the inflationary e-folding number $N_{\rm I}$, scalar spectral index, $n_s$ and tensor-to-scalar ratio, $r$. If one assumes that after the reheating phase, the comoving entropy density remains conserved, then such conservation law imposes an interesting relation among the parameters ($N_{\rm I},\,N_{\rm re},\,T_{\rm re}$) as follows \cite{Dai:2014jja,Cook:2015vqa}, 
\begin{eqnarray} \label{entc}
    T_{\rm re}&=&\left(\frac{43}{11\,g_{\star\rm re}}\right)^{1/3}\,\left(\frac{H_{\rm I}}{k_\star/a_{0}}\right)\,e^{-(N_{\rm I} +N_{\rm re})}\,T_0\,\nonumber \\
 &=& 2.5\times 10^{39} \left(\frac{H_{\rm I}}{10^{13}~\mbox{GeV}}\right)\,e^{-(N_{\rm I} +N_{\rm re})} \,\mbox{GeV}
\end{eqnarray}
Where, $k_\star/a_0=0.05~ \mbox{Mpc}^{-1}$ as the CMB pivot scale, the present CMB temperature $T_0=2.725$ K and $a_0$ is the present day scale factor. $H_{\rm I}$ is the inflationary Hubble parameter and $N_{\rm re}$ be the e-folding number during reheating.
Combining the  Eq.\ref{entc} and Eq.\ref{Trephi} (and \ref{TreN}), in Fig. \ref{nsr}, we scan the $n_s$-$r$ parameter space for different values of $\alpha$ for $w_\phi=0.50$ (top) and  $w_\phi=9/11$ (bottom). The black-dot correspond to $T_{\rm BBN}\,(T^{\rm GW}_{\rm re})$ point of $\wphi=1/2\,(9/11)$  for the neutrino heating case. Whereas green, red and blue dots correspond to the $T_{\rm BBN}$ for neutrino dominating case for three different masses of $M_3 = (M^{\rm min}_3\,,5\times 10^{11}, 2\times 10^{13})$ GeV respectively. The segments between the black dot and any other color dot associated with a particular mass of $M_3$ are the allowed $n_{\rm s}$ range.
 \begin{figure}[t]
\centering
\includegraphics[width=\columnwidth]{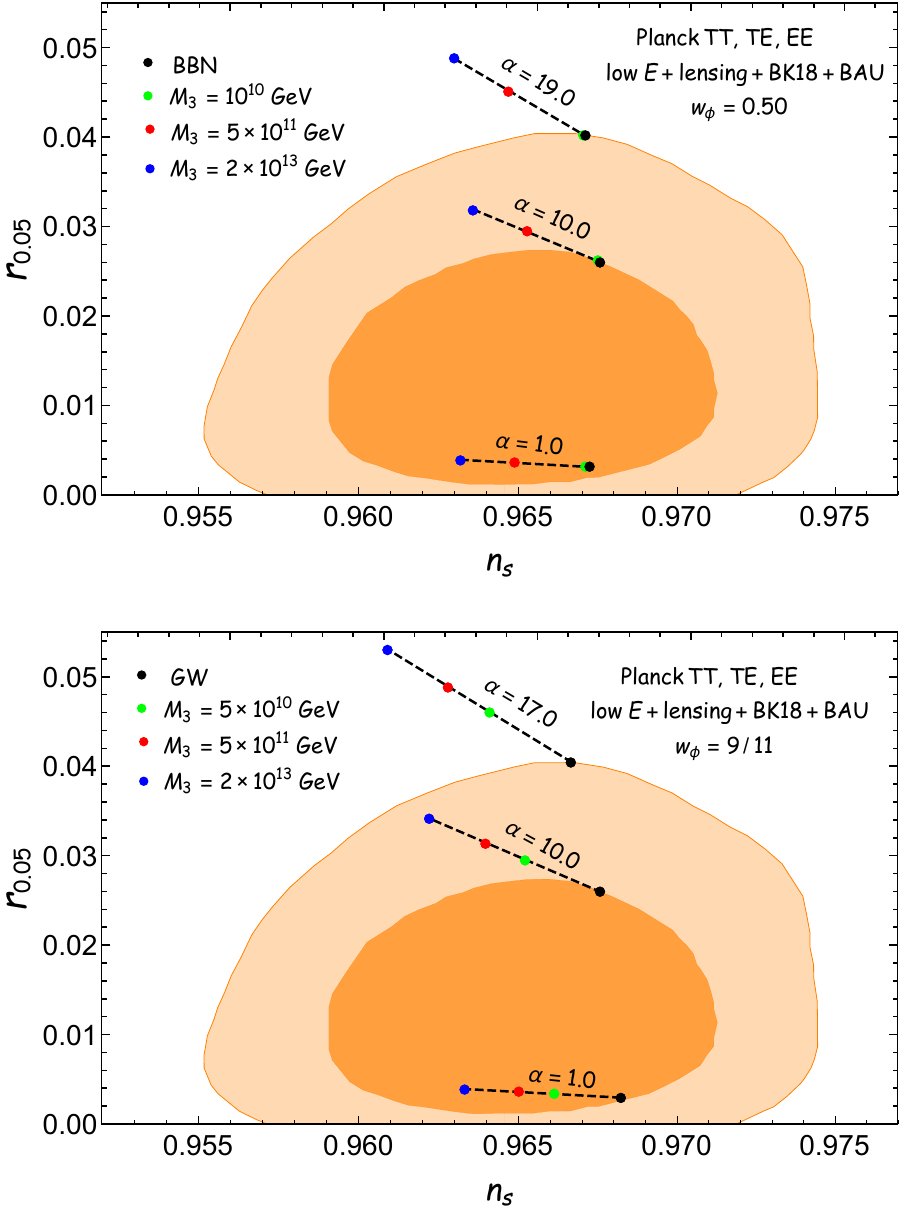}
\caption{ Compare our
result with the observational $68\%$ and $95\%$ CL constraints from BICEP/Keck, in the $(ns, r)$ plane for $\wphi=0.50$ (top) and $\wphi=9/11$ (bottom). The black-dot points correspond to $T_{\rm BBN}\,(T^{\rm GW}_{\rm re})$ point of $\wphi=0.50\,(9/11)$  for the neutrino heating case, and the different color points correspond to the $T_{\rm BBN}$ points for different mass $M_3$ for the neutrino dominating case.}
\label{nsr}
\end{figure}
For example, if one considers $M_3=M^{\rm min}_3\simeq5\times10^{10}$ GeV, $w_{\phi}=9/11$ and $\alpha =1$, the allowed range of $n_s$ will lie within $0.96615 \leq n_s \leq 0.96825$ (segment between black dots and green dots)  which is well inside the $1\,\sigma$ bound. Consequently, the inflationary e-folding number within $59\leq N_{\rm I}\leq 64$. Interestingly, for $\alpha=1.0$, the whole ranges of $n_s$ lie within $1\sigma$ regions. For $w_\phi=1/2\,(9/11)$, the prediction of $n_s$ lying within $0.96329\,(0.96321) \leq n_s \leq 0.96825\,(0.96724)$.
Similar bounds can also be obtained for $\alpha =10, 17,19$, and that can be decoded from the Fig.\ref{nsr}.

\section{Leptogenesis and constraints}{\label{sc5}}
RHNs are gravitationally produced from the inflaton. Therefore, all of them undergo CP-violating out-of-equilibrium decay and produce lepton asymmetry. By the well-known non-perturbative sphaleron processes, those lepton asymmetries are then converted into the baryon asymmetry. For our analysis we  considered the following mass hierarchy $M_{1}\lesssim m^{\rm end}_{\rm\phi}\ll M_2$. Note that due to its very small decay width, $\nu_{\rm R}^3$ does not contribute significant asymmetry. The lepton asymmetry will, therefore, predominantly be produced by the decays of $\nu^1_{\rm R}$ \footnote{Note that the three-body decay such as $\nu^{1,2}_R\rightarrow\nu^3_R+\nu_L+\nu_L$ is kinematically allowed. However, we checked that the associated production rate of $\nu^3_R$, $\Gamma_{1\rightarrow3}\simeq\frac{\beta^2\,(y^\dagger\,y)_{11/22}}{2\,M_{1/2}}\left(\frac{3}{2}p^2+\frac{1}{2}-2\,p-p^2\,\ln{p}\right)\,\,,p=M^2_3/M^2_{1/2}$) is  very much  suppressed compared to the gravitational contributions.}. The  CP asymmetry parameter $(\epsilon_{\Delta L})$ generated from the decay of $\nu^1_{\rm R}$, using the seesaw mechanism  \cite{Buchmuller:2004nz,Barman:2022qgt,Kaneta:2019yjn,Barman:2023opy} is expressed as , 
\begin{equation}
\begin{aligned}
\epsilon_{\Delta L}
&=\frac{\sum_j[\Gamma(N_1\rightarrow l_j H)-\Gamma(N_1\rightarrow \bar l_j H^\star)]}{\sum_j[\Gamma(N_1\rightarrow l_j H)+\Gamma(N_1\rightarrow \bar l_j H^\star)]} \\
&\simeq \frac{3\,\delta_{\rm eff}}{16\,\pi}\frac{m_{\rm\nu,max}\,M_{1} }{v^2}\,\\
&\simeq9.85\times10^{-5}\left(\frac{M_1}{10^{12}\,\mbox{GeV}}\right)
\end{aligned}
\end{equation}
The final lepton asymmetry depends on the out-of-equilibrium number density of $\nu^1_{\rm R}$ and the CP asymmetry $(\epsilon_{\Delta L})$. 
\begin{figure*}
          \begin{center}
\includegraphics[width=16.50cm,height=6.0cm]{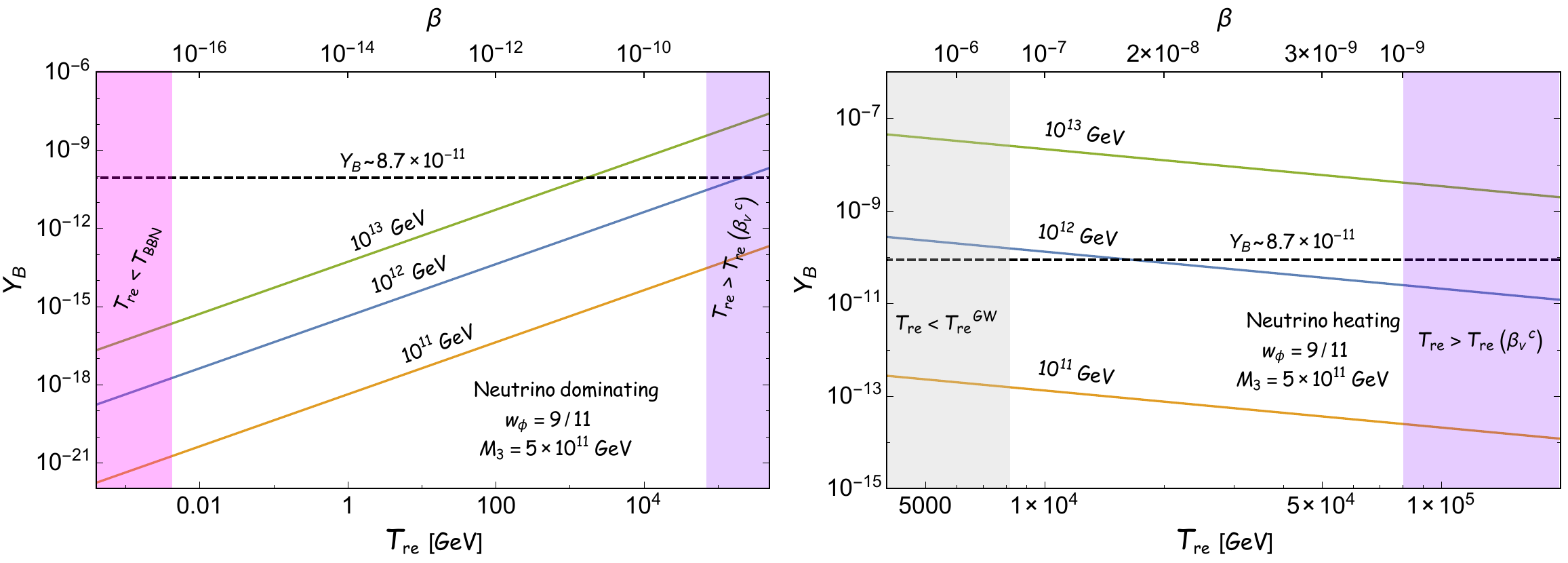}
         \caption{$Y_{\rm B}$ as function of $\tre$ for three different choices of $M_{1}$ (different solid line) for a fixed $M_{3}=5\times10^{11}$ GeV and $\wphi=9/11$. The magenta-shaded and gray-shaded region corresponds to $\tre < T_{\rm BBN}$ and $\tre < T_{\rm re}^{\rm GW}$ respectively. The purple-shaded region is correspond to $\tre>T_{\rm re}(\beta^{\rm c}_{\nu})$. The horizontal dashed line corresponds to the observed baryon asymmetry $Y_{\rm B}\sim 8.7\times10^{-11}$ GeV }
 \label{ylepto}
          \end{center}
      \end{figure*}
Where, $\delta_{\rm eff}$ is the
effective CP-violating phase in the neutrino mass matrix, which typically assumes the value within $0\leq\delta_{\rm eff}\leq1$, and we have taken $\delta_{\rm eff}=1$. 
We take the normal hierarchy of active neutrino mass with $m_{\rm\nu,max}=m_3=0.05$ eV \footnote{For the inverted hierarchy $m_{\rm\nu,max}=m_2=0.05$ eV.}, and the Higgs vacuum expectation value $v =174$ GeV.
As stated earlier, the above lepton asymmetry is converted into the baryon asymmetry $ (Y_{\rm B})$ via the electroweak sphaleron processes \cite{sy,ja},
\begin{equation}{\label{yb}}
    Y_{\rm B}=\frac{n_{\rm B}}{s}=a_{\rm sph}\,\epsilon_{\Delta L}\frac{n_{1}(T_{\rm re})}{s(T_{\rm re})}\,,
\end{equation}
where $a_{\rm sph}=\frac{28}{79}$ and $s(T_{\rm re})=\frac{2\pi^2}{45}g_{\rm\star s}T^3_{\rm re}=\frac{4\,\epsilon}{3}\frac{g_{\rm\star s}}{g_{\rm\star r}}T^3_{\rm re}$ is the entropy density at the end of reheating. Where $g_{\rm \star s}$ counts the number of entropic degrees of freedom. The number density, $n_{1}(T_{\rm re})$ at the time of reheating from Eq.\ref{Nis}, 
\begin{equation}\label{Nisre}
    n_{\rm 1}(\tre)= \frac{3H^3_{\rm end}M^2_{\rm 1}\Sigma_{\rm 1}}{2\pi(1-w_\phi)(\gamma \,m^{\rm end}_\phi)^2} \left(\frac{\are}{\aend}\right)^{-3}\,,
\end{equation}
We will now use this expression in Eq.\ref{yb} to constrain our model parameters for two different reheating cases. 
\subsection{Neutrino dominating: $\beta \leq \beta^{\rm c}_\nu$ }
  Combining the Eqs.(\ref{areN}) and (\ref{TreN}), one can find the following relation between $\are$ and $\tre$,
\begin{equation}{\label{are2}}
   \frac{a_{re}}{a_{\rm end}}=\left(\frac{H_{\rm end}}{4.7\,M_{\rm p}}\right)^{2/3}\left(\frac{M_{\rm p}}{T_{\rm re}}\right)^{4/3}\left(\frac{a_{\rm\nu}}{a_{\rm end}}\right)^{-w_{\rm\phi}}
\end{equation}
Utilizing the Eqs. \ref{yb}, \ref{Nisre} and \ref{are2}, $Y_{B}$ can be written as in terms of $\tre$, 
\begin{equation}{\label{Ybfinal1}}
   \begin{aligned}
        Y_{\rm B}&=\frac{21}{79}\frac{\epsilon_{\rm\Delta L}}{\epsilon}\frac{g_{\star s}}{g_{\star r}}\frac{2\epsilon\, n^{\rm end}_{\rm 1}}{3M^3_{\rm p}}\left(\frac{M_{\rm p}}{H_{\rm end}}\right)^{2}\frac{T_{\rm re}}{M_{\rm p}}\left(\frac{a_{\rm\nu}}{a_{\rm end}}\right)^{3w_{\rm\phi}},\\
       &\simeq4\times10^{-15}\frac{\Sigma_1}{\Sigma_3}\left(\frac{M_1}{10^{12}\,\mbox{GeV}}\right)^3\left(\frac{5\times10^{11}\,\mbox{GeV}}{M_3}\right)^3\\
       &\quad\quad\times\left(\frac{T_{\rm re}}{4\,\mbox{MeV}}\right)
   \end{aligned}
\end{equation}
We have also used Eq.\ref{AphiN} to find the above expression, and we assume $g_{\rm \star r} = g_{\rm \star s} \simeq  100$. In this scenario, the final baryon asymmetry is independent of $w_\phi$ but depends on the reheating temperature $\tre$. From the above equation, for a given $M_3$, $Y_{\rm B}\propto M_1^3\,T_{\rm re}\propto M_1^3\,\beta$, the
 baryon asymmetry increases with increasing both $\tre$ (or $\beta$) and $M_1$.  We demonstrate this behavior in Fig.\ref{ylepto} (left plot), where $Y_{\rm B}$ is plotted as a function of $\tre$ for different $M_1$. Therefore, to find the correct baryon asymmetry, we need to decrease $M_1$ for increasing $\beta$ (see dashed line in  Fig.\ref{pre}). Fixing the observed baryonic asymmetry, therefore, leads to $M_1 \propto \beta^{-1/3}$. This slope is also recovered numerically as shown by the dashed line in the upper part of the right Fig.\ref{pre} (dashed line) for $w_\phi = 9/11$. 

The required $\tre$ for producing the observed baryon asymmetry is not achievable for $\wphi<0.50$. For example, $M_1=10^{12}$ GeV, $M_3=10^{10}$ GeV, we need $\tre\sim 1$ GeV to satisfy the correct asymmetry. But using these $M_3$, $\tre\sim 1$ GeV is not achievable $w_\phi\leq0.5$. For this reason, we never get a baryon asymmetry universe for  $w_{\phi}=1/3\,,0.5$. However, for $w_{\phi}>0.50$, for reasonable $T_{\rm re}$ and $M_1$, the proper asymmetry can be generated. For example, for $\wphi=9/11$,  $M_3=5\times10^{11}$ GeV, we have found the correct asymmetry has been generated when $\tre\geq 100$ GeV (see left upper plot in Fig.\ref{pre}).
\subsection{Neutrino heating: $ \beta^c_\nu \leq \beta \leq   \beta^{\rm c}_\phi$} 
For this case, using the Eqs.\ref{yb}, \ref{Nisre} and \ref{are1}, $Y_{B}$ can be written as 
\begin{equation}{\label{ybphi}}
\begin{aligned}
    Y_{B}(T_{re})&=\frac{21}{79}\frac{\epsilon_{\rm\Delta L}}{\epsilon}\frac{g_{\star s}}{g_{\star r}}\frac{n^{\rm end}_{\rm1}}{M^3_{\rm p}}\left(\frac{\sqrt\epsilon M_{\rm p}}{\sqrt3H_{\rm end}}\right)^{\frac{2}{1+w_{\rm\phi}}}\left(\frac{T_{\rm re}}{M_{\rm p}}\right)^{\frac{1-3w_{\rm\phi}}{1+w_{\rm\phi}}},\\
    &=4\times10^{\frac{30w_\phi-32}{1+w_\phi}}\frac{\Sigma_1}{(1-w_\phi)\gamma^2}\left(\frac{M_1}{10^{12}\,\mbox{GeV}}\right)^3\\
    &\times\left(\frac{10^{13}\,\mbox{GeV}}{m^{\rm end}_\phi}\right)^2\left(\frac{H_{\rm end}}{ 10^{13}\mbox{GeV}}\right)^{\frac{1+3w_\phi}{1+w_{\rm\phi}}}\left(\frac{T_{\rm re}}{\mbox{GeV}}\right)^{\frac{1-3w_{\rm\phi}}{1+w_{\rm\phi}}}
    \end{aligned}
\end{equation}
 
In this scenario, the final baryon asymmetry is dependent on both $w_\phi$ and  $\tre$ and behaves as $Y_{\rm B}\propto M^3_1\,T_{\rm re}^{\frac{1-3w_{\rm\phi}}{1+w_{\rm\phi}}}\propto M^3_1\beta^{1/(1+w_{\rm\phi})}$, for a given $M_3$. As reheating happens for $\wphi>1/3$,  the
 baryon asymmetry decreases (increases) with increasing $\tre\,(\beta)$. We confirmed this behavior in Fig.\ref{ylepto} (right plot). Fixing the observed baryonic asymmetry leads $M_1 \propto \beta^{-1/(3+3w_{\rm\phi})}$. For example, considering $w_\phi = 9/11 (1/2)$, the corresponding slope turns out to be
$M_1 \propto \beta^{-11/60} (\beta^{-2/9})$.
This slope also recovered numerically, as shown in the upper part of Fig.\ref{pre}.
\begin{figure*}
          \begin{center}
\includegraphics[width=16.50cm,height=5.50cm]{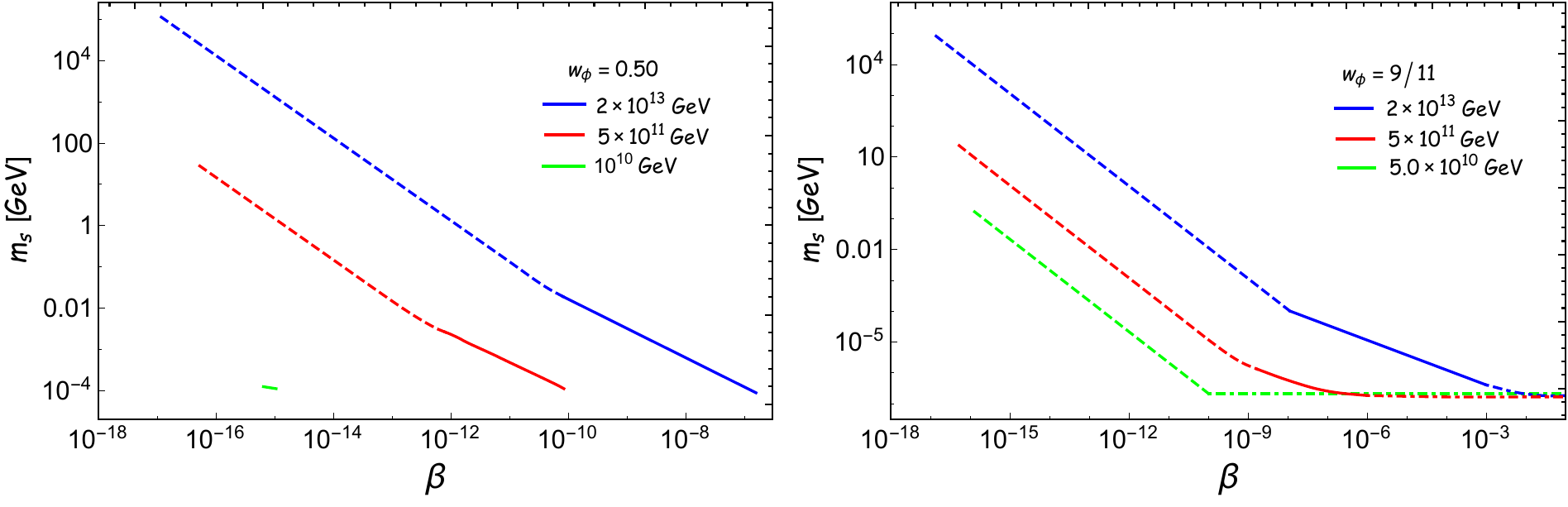}
         \caption{We have shown the relation between the $\beta$
and the mass of DM $\ms$, that predicts the observed DM
relic abundance from purely gravitational interactions. The different color lines correspond to different  $M_3$. The solid (dashed) lines correspond to the neutrino heating (dominating) case. The dot-dashed lines are excluded by an excess of PGWs.}
 \label{gradm}
          \end{center}
      \end{figure*}
For $w_{\phi}=1/3$, the produce asymmetry is too small that we never get a baryon asymmetry universe. However, for $w_{\phi}\geq0.5$, for reasonable $T_{\rm re}$ and $M_1$, the proper asymmetry can be generated. For $w_{\phi}=1/2$, the proper asymmetry will be generated when $T_{\rm re}$ is close to BBN temperature. For $w_{\phi}=1/2$, $T_{\rm re}$ range where proper asymmetry will be generated is $4$ MeV to $10$ MeV, the corresponding mass range $M_1\simeq(1.2-2.0)\times10^{13}$ GeV. If we increase $w_\phi$, the relaxation of $T_{\rm re}$ and $M_1$ will occur. For example $w_{\rm\phi}=9/11$ and $M_3=5\times10^{11}$ GeV, mass range $M_1\simeq(8-20)\times10^{11}$ GeV.

\section{Particle Dark matter phenomenology}{\label{sc6}}
In the recent past, the effect of non-standard cosmology has gained significant interest in the DM phenomenological studies \cite{Maity:2018dgy,Garcia:2020eof,Garcia:2020wiy,Garcia:2021gsy,Maity:2018exj,Haque:2020zco,Giudice:2000ex,Barman:2022tzk,Bhattiprolu:2022sdd,Harigaya:2014waa,Harigaya:2019tzu,Okada:2021uqk,Ghosh:2022fws,Haque:2021mab,Ahmed:2022tfm,Bernal:2022wck,Bernal:2023ura,Chowdhury:2023jft}. In this paper, we aim to discuss how the present neutrino reheating scenario affects the DM production and constrain its parameter space. We consider both thermal and non-thermal DM production scenarios. Let us consider a real standard model singlet DM $S$ \cite{Silveira:1985rk,Burgess:2000yq,McDonald:1993ex,Andreas:2008xy,Davoudiasl:2004be,Vangsnes:2021rni,Lebedev:2021xey,Guo:2010hq,Yaguna:2011ei,Cline:2013gha}, which is odd and all the standard model particles are even under the $\mathbb Z_2$ symmetry. The  DM interacts with the SM via the Higgs portal interaction and also with inflaton through universal gravitational interaction. The interaction Lagrangian for DM is,
\begin{equation}
    \mathcal{L_{\rm int}}=\frac{h_{\rm\mu\nu}}{M_{\rm p}}T^{\rm S}_{\rm\mu\nu}+\lambda_{\rm hs}S^2H^\dagger H,
\end{equation}
where $h_{\mu\nu}$ is the graviton field, $T^{\rm S}_{\rm\mu\nu}$ is the energy-momentum tensor of DM and $\lambda_{\rm hs}$ is the Higgs portal coupling.\\
\begin{figure*}
 	\begin{center}
\includegraphics[width=15.0cm,height=3.0cm]{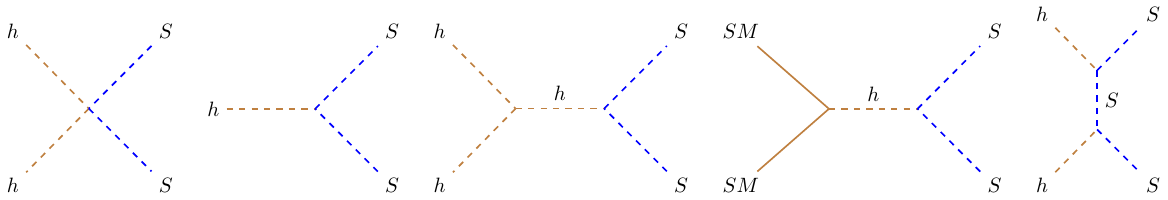}\quad
\caption{Leading diagrams for Higgs portal dark matter production}\label{fynmanrad}
 	\end{center}
 \end{figure*}

\begin{figure*}
          \begin{center}
\includegraphics[width=17.50cm,height=12.0cm]{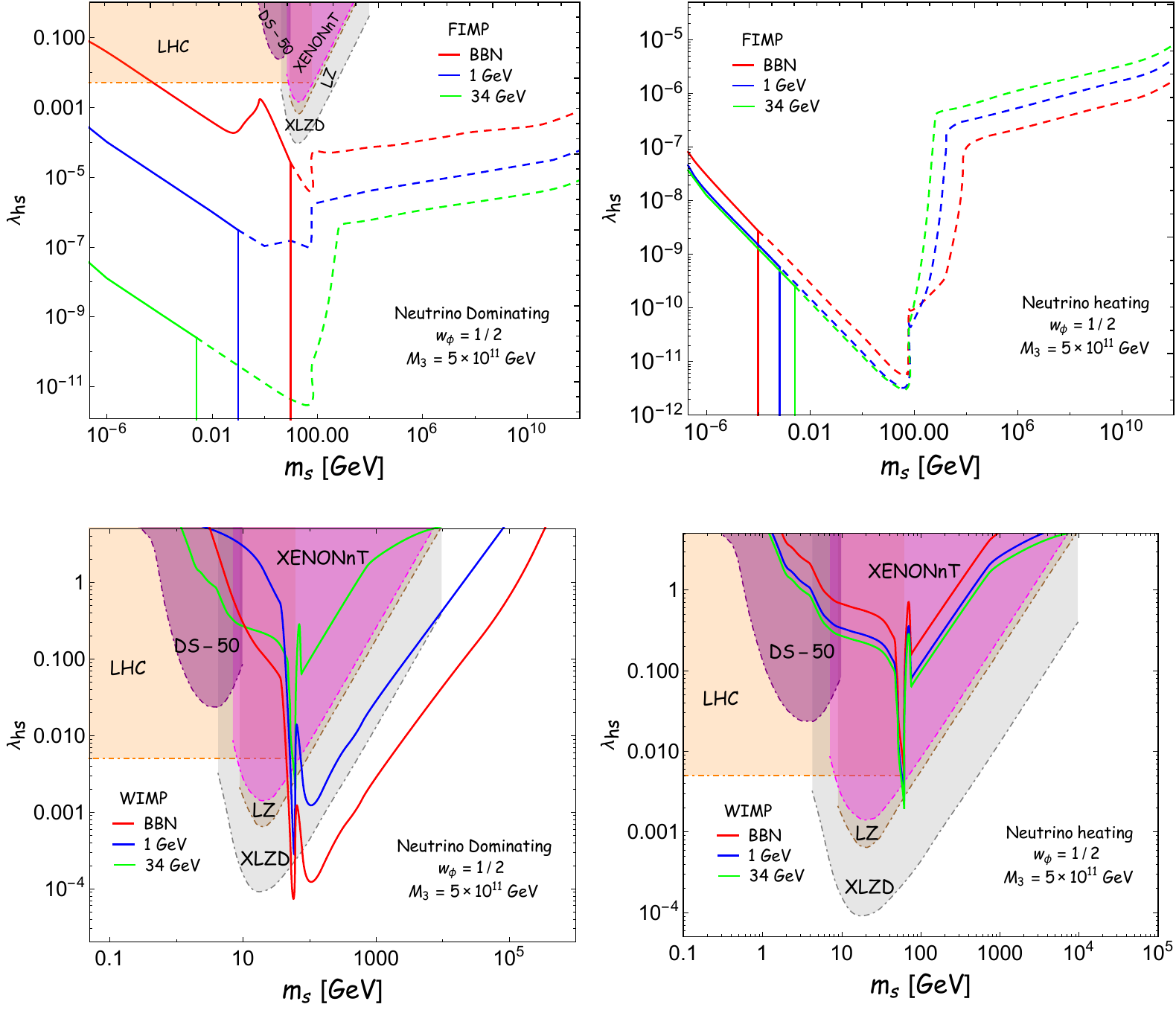}
         \caption{ The predicted Higgs portal coupling $\lmd$ as a function of the DM mass $\ms$ from the observed DM abundance for $\wphi=1/2$ for both neutrino dominating (left-column) and neutrino heating (right-column). The different color lines correspond to different reheating temperatures $\tre$, where restricts $\tre$ within the allowed ranges (minimum $\tre=T_{\rm BBN}$ to maximum $\tre=34 ~\rm GeV$). For FIMP (top plot), the dashed part of the lines is excluded due to overproduction from the
gravitational contribution. The different color-shaded regions represent the experimental constraints and projections.}
 \label{SSD2}
 \end{center}
\end{figure*}
\textbf{Gravitational DM production from inflaton : }
Gravitational production of DM from the background inflaton field is universal and hence will always be present in any
inflationary scenario. Since we have considered scalar DM, the  production rate is,
\begin{equation}
    R^{\rm \phi}_{\rm s}=
         \frac{\rho^2_{\rm \phi}}{8\pi M^4_{\rm p}}\Sigma^{0}_{\rm s}\,,
\end{equation}
where, 
\begin{equation}{\label{rx}}
     \Sigma^{0}_{\rm s}= \sum^{\infty}_{\nu=1}\lvert\mathcal P^{2n}_\nu \rvert^2\left(1+\frac{2m^2_{\rm s}}{\nu^2\gamma^2m^2_{\rm\phi}}\right)^{2}\left(1-\frac{4m^2_{\rm s}}{\nu^2\gamma^2m^2_{\rm\phi}}\right)^{1/2}\,.
\end{equation}
 accounts for the sum over the Fourier modes of the inflaton potential, and its numerical values are given in Table-\ref{fouriersum}, and $m_{\rm s}$ is the DM mass.
The evolution of DM number density ($n_{\rm s}$) is governed by the Boltzmann equation
\be
\dot{n}_{\rm s}+3H\,n_{\rm s}=R^{\rm s}_{\rm\phi}\,.
\label{numx}
\ee
Using the Eq. \ref{rx} in Eq.\ref{numx}, we have obtained the following solutions for the number
density
\begin{equation}
  n_{\rm s}(a)\simeq  \frac{3H^3_{end}\Sigma^{0}_{\rm s}}{4\pi(1+3w_\phi)}(a/a_{\rm end})^{-3} .
\end{equation}
The present-day DM relic abundance can be written as
\begin{eqnarray}
    \Omega_{\rm s}h^2&=&\Omega_{\rm R}h^2\frac{m_{\rm s}\,n_{\rm s}(a_{\rm re})}{\epsilon\, T_{\rm now}\,T^3_{\rm re}} \nonumber\\
   &=&  \Omega_{\rm R}h^2 \frac{3\,\Sigma^{0}_{\rm s}}{4\,\pi\epsilon(1+3w_\phi)}\frac{H^3_{end}m_{{\rm s}}}{T_{\rm now}T^3_{\rm re}}\left(\frac{a_{\rm re}}{a_{\rm end}}\right)^3 .
\end{eqnarray}
Note that the final abundance of the gravitational DM is dependent not only on the reheating temperature but also on the duration of reheating $a_{\rm re}/a_{\rm end}$. Utilizing $a_{\rm re}/a_{\rm end}$ from Eq.\ref{are2},  
the DM abundance for neutrino dominating case turns out to be, 
\begin{equation}{\label{dma}}
   \begin{aligned}
        \Omega_{\rm s}h^2 &\simeq0.12\frac{\Sigma^0_s}{\Sigma_3}\left(\frac{m_{\rm s}}{\mbox{200\,GeV}}\right)\left(\frac{5\times10^{11}\,\mbox{GeV}}{M_3}\right)^3\\
       &\quad\quad\times\left(\frac{m^{\rm end}_\phi}{10^{13}\,\mbox{GeV}}\right)^2\left(\frac{T_{\rm re}}{4\,\mbox{MeV}}\right),
   \end{aligned}
\end{equation}
and similarly utilizing $a_{\rm re}/a_{\rm end}$ from Eq.\ref{are1} for the neutrino heating case, we have the following expression for the DM abundance,
\begin{equation}{\label{abundance1}}
\begin{aligned}
        \Omega_{\rm s}h^2\simeq&10^{\frac{4(11\wphi-4)}{1+\wphi}}\frac{\Sigma^{0}_{\rm s}}{1+3w_\phi}\left(\frac{m_{\rm s}}{\mbox{GeV}}\right)\\
        &\quad\times\left(\frac{\hend}{10^{13}\,\mbox{GeV}}\right)^{\frac{1+3\wphi}{1+w_{\rm\phi}}}\left(\frac{T_{\rm re}}{\mbox{GeV}}\right)^{\frac{1-3w_{\rm\phi}}{1+w_{\rm\phi}}} .
        \end{aligned}
\end{equation}
Due to their distinct reheating histories, one notes DM abundance behaving as 
$\Omega_{\rm s}h^2\propto \tre \propto \beta$ for neutrino dominating, and  $\Omega_{\rm s}h^2\propto \tre^{\frac{1-3w_\phi}{1+w_\phi}} \propto\,\beta^{1/(1+\wphi)}$for neutrino heating case. Once the observed dark matter (DM) abundance is satisfied, these two behaviors are reflected in the sudden change of the slope in the $m_{\rm s}$ vs. $\beta$ curve, as shown in Fig.\ref{gradm}. 
This clearly depicts how the neutrino sector such as coupling parameter $\beta$ is correlated with the DM sector such as mass $m_s$. The above production is purely gravitational and non-thermal in nature. In the following section, we discuss how neutrino reheating also controls DM production when DM-Higgs portal coupling is taken into account.
 \begin{figure*}
          \begin{center}
\includegraphics[width=17.0cm,height=12.0cm]{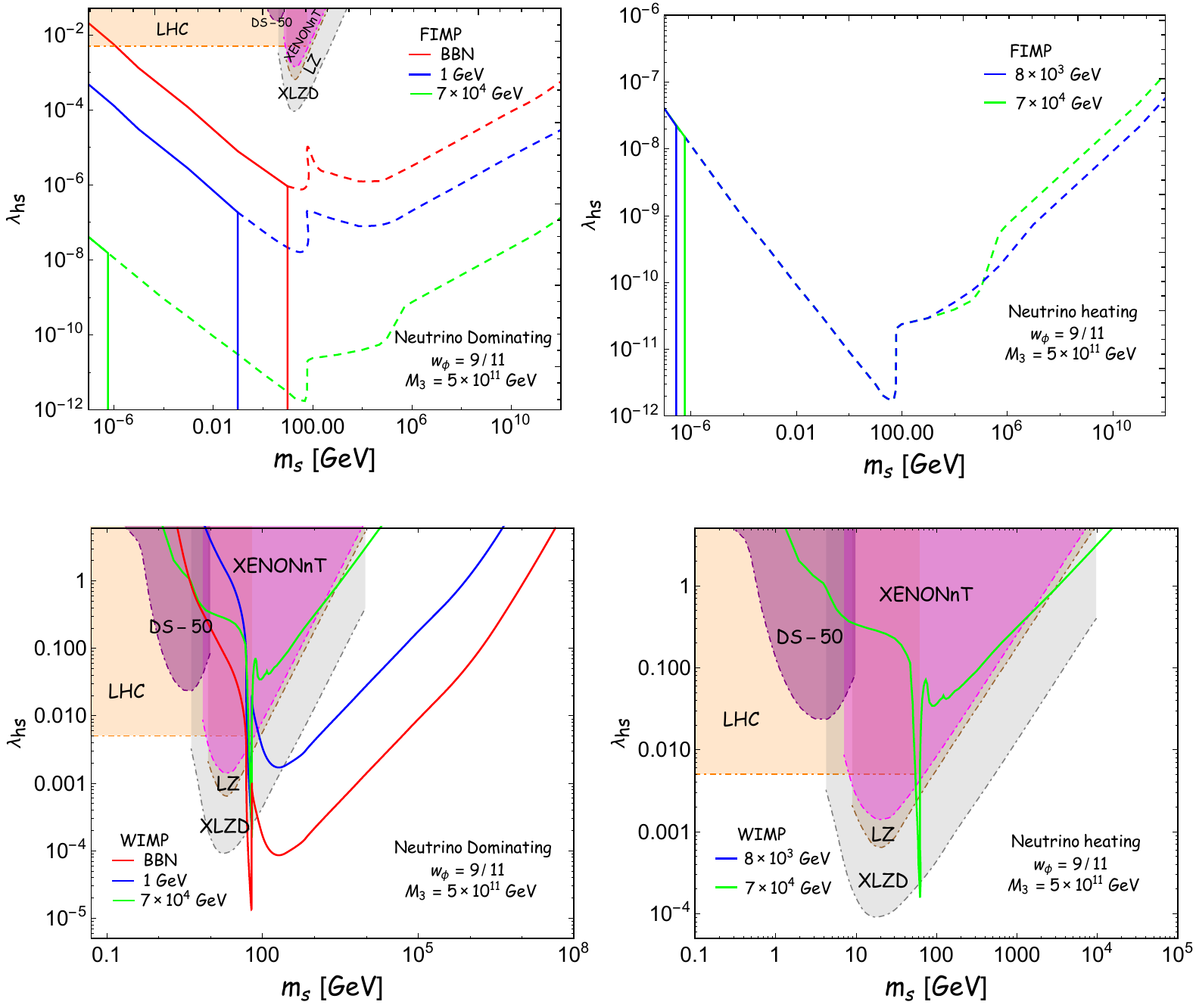}
         \caption{The predicted Higgs portal coupling $\lmd$ as a function of the DM mass $\ms$ from the observed DM abundance for $\wphi=9/11$. The different color lines correspond to different reheating temperatures $\tre$, where restricts $\tre$ within the allowed ranges, minimum $\tre=T_{\rm BBN}\,(\tre^{\rm GW})$ to maximum $\tre=7\times10^{4} ~\rm GeV$ for neutrino dominating (heating). Detailed descriptions are the same as Fig.\ref{SSD2}.}
 \label{SSD3}
 \end{center}
      \end{figure*}     
 
\textbf{Higgs portal DM :} 
In the Higgs sector, the electroweak scale is an important energy scale where the phase transition occurs, and all the standard model particles acquire mass. Before the electroweak symmetry breaking (EWSB), i.e.,$T>T_{\rm EW}\simeq160$ GeV, DM particles are produced only through Higgs annihilation via contact diagram $HH\rightarrow SS$. After EWSB i.e $T<T_{\rm EW}$, the Higgs boson acquires VEV($v$) and expanding the contact operator $S^2H^\dagger H$ around v as $H= v + h$, we have
\begin{equation}
    \frac{\lambda_{hs}}{2}(h^2+2vh)S^2+...\,.
\end{equation}
The first term represents the usual four-point interaction, and the last term generates the $3$-point vertex $h\rightarrow SS$, which not only accounts for the direct Higgs decay but also enables massive SM particles to annihilate into DM through $s$-channel Higgs exchange, $S$- mediated $t$- and $u$- channel (see Fig.\ref{fynmanrad}). In
addition, the universal gravitational production of DM will always be present, which can
not be ignored. Thus, the Boltzmann equations associated with
DM takes the following form
\begin{equation} {\label{darkd1}}
\dot n_{\rm s}+3Hn_{\rm s}+\langle\sigma v\rangle( n_{\rm s}^2- n_{\rm eq}^2)+\langle\Gamma_{h}\rangle n_{\rm h} \left(1-\frac{n^2_{\rm s}}{n^2_{\rm eq}}\right)-R^{\rm \phi}_{\rm s}=0\,,
\end{equation}
where $n_{\rm eq}\,(n_{\rm h})$ is the  equilibrium DM (Higgs) number density and $\langle\Gamma_{h}\rangle$ is the thermally average $h\rightarrow SS$ decay width which is
\begin{equation}
\langle\Gamma_{h}\rangle=\Gamma_{\rm h\rightarrow SS}\frac{K_{1}\left(\frac{m_{h}}{T}\right)}{K_{2}\left(\frac{m_{h}}{T}\right)},\left[\Gamma_{\rm h\rightarrow SS}=\frac{\lambda_{\rm hs}^2v^2}{8\pi m_{h}}\sqrt{1-\frac{4m_{\rm s}^2}{m^2_{h}}}\right]
\end{equation}
where $K_{n}$ is the modified Bessel function of the $n$-th order and $m_{h}=125$ GeV is the Higgs mass. $\langle\sigma v\rangle$ is the thermally averaged annihilation cross-section \cite{Gondolo,Guo:2010hq,Bernal:2018kcw} of DM for all the relevant processes discussed before (see the Appendix \ref{sigmav} where we listed the cross-section for all the process). Depending on the strength of the portal coupling $\lmd$, DM particle production $S$ can either be thermally or non-thermally produced. For the well-known freeze-out mechanism for the weakly interacting massive particle(WIMP), the portal coupling typically has to be much larger compared to the freeze-in mechanism for feebly interacting massive particles (FIMP). The required value of the portal coupling $\lmd$ to achieve the observed DM abundance is plotted as a function of DM mass $m_{\rm s}$ in Fig.\ref{SSD2} and \ref{SSD3} for both FIMP (top plot) and WIMP (bottom plot) type DM. The different color lines correspond to different $\tre$. In the region above (below) the $\lmd-m_{\rm s}$ lines, the predicted relic density is overproduced (underproduced) for FIMP, and for WIMP, it is the opposite. we have further shown the DM parameter space ($\lmd\,,m_{\rm s}$) with current and future projected experimental bounds (shown in shaded regions). Note that for the FIMP-like production in Fig.\ref{SSD2}, \ref{SSD3}, the dashed part of the lines is excluded due to overproduction from the gravitational contribution. The upper bound of FIMP-like DM mass can be determined from the Eq. \ref{dma}, \ref{abundance1} after satisfying observed DM abundance and can be decoded from the solid vertical lines for different $\tre$ values from the upper two panels of the Fig. \ref{gradm}. As indeed can be seen, if one considers $\wphi=1/2$ and $M_3=5\times10^{11}$ GeV, for $\tre=4$ MeV, the upper bound of $\ms$ is $10$ ($10^{-4}$) GeV for neutrino dominating (neutrino heating) case. A similar bound can also be decoded for $w_\phi = 9/11$ from Fig. \ref{SSD3}. For the FIMPs production, the viable mass range for the neutrino-dominating case turned out to be within $10^2\,\mbox{eV}\leq m_s\leq10\,
  \mbox{GeV}$ for both $\wphi=1/2\,,9/11$. Similarly, for the neutrino-heating case, the viable mass range turned out to be within $10^2\,\mbox{eV}\leq \ms\leq10^{-3}\,(10^{-6})$ GeV for $\wphi=1/2\,(9/11)$. Note that any values lower than the bound of $100\,\mbox{eV}$ FIMP mass will give the under abundance today due to extremely weak production, and this bound is independent of reheating histories \cite{Haque:2023yra}.

We performed our analysis restricting the coupling parameter $\lmd\leq\sqrt{4\pi}$ \cite{Han:2015hda} in the perturbative limit, and that automatically limits the possible DM mass range as illustrated in Fig.
\ref{SSD2}, \ref{SSD3}. For WIMPs, the mass range turned out to be within $1\,\leq m_s\leq10^{5.6}\,(10^{7.7}) \,\mbox{GeV}$ for neutrino dominating with $\wphi=1/2\,(9/11)$ and within $1\,\leq m_s\leq 10^4 \,\mbox{GeV}$ for both $\wphi =1/2\,,9/11$ in the neutrino heating case, while keeping fixed value of the neutrino mass $M_3=5\times10^{11}\,$GeV. The reheating phase can affect the DM relic if the DM production is completed before the end of reheating, which happens for $m_s\gtrsim 25\,\tre$ \cite{Bernal:2022wck,Haque:2023yra,Silva-Malpartida:2023yks} for WIMP like particles.
For FIMPs, it is $m_s\geq\tre$ if $\tre$ is larger than $m_h/2$, but if $\tre$ is lower than $m_h/2$, all DM particles can freeze in during reheating. When the DM production is completed before reheating ends, the DM relic suffers a high entropy dilution, which demands larger DM production to achieve the current DM relic. Thus, FIMPs (WIMPs) require larger (smaller) coupling to satisfy the observed DM abundance.

In Fig. \ref{SSD2} and \ref{SSD3}, we projected the DM parameter space ($\lmd\,,m_s$) in two distinct reheating background with reference to various current and future projected experimental constraints. In the context of Higgs portal coupling, the Higgs decaying into DM is severely constrained by the Large Hadron Collider (LHC) experiment. For instance if the DM mas satisfies $m_s<m_h/2$, the decay $h\rightarrow\,SS$ is naturally kinematically allowed and contributes to the invisible Higgs decay width  $\Gamma_{h\rightarrow\rm SS}$. The corresponding invisible branching ratio ($\rm{BR}_{\rm inv}=\Gamma_{h\rightarrow\rm SS}/(\Gamma_h+\Gamma_{h\rightarrow\rm SS})$) is constrained by the LHC searches to be $\rm{BR}_{\rm inv} \lesssim 0.11$ at $95\%\,$CL \cite{ATLAS:2022yvh,CMS:2023sdw,ATLAS:2023tkt}. Using the visible Higgs decay with $\Gamma_{\rm h}\simeq4.07$ MeV \cite{Djouadi:2005gi,CMS:2023sdw}, the bound on the portal coupling for $m_s<m_h/2$ is $\lmd\lesssim5\times10^{-3}$ as depicted in orange region ((labeled as "LHC") in the figures \ref{SSD2} and \ref{SSD3}. Dark matter direct detection experiments, on the other hand, impose further constraints on the model parameters. The effective spin-independent cross-section $(\sigma_{\rm SI})$ for elastic scattering between DM and nucleons is determined by the equation
\begin{equation}
    \sigma_{\rm SI}=\frac{\lmd^2\,\mu_{\rm N}^2\,f_{\rm N}^2\,m_{\rm N}^2}{4\pi\,m_s^2\,m_h^4}\,,
\end{equation}
$\mu_{\rm N}=m_{\rm N}\,\ms/(m_{\rm N}+\ms)$ is the reduced mass of the DM-nucleon system with nucleon mass $m_{\rm N}=1\,$GeV and $f_{\rm N}=0.30$. Taking  $\sigma_{\rm SI}$ from various direct-detection experiments such as DarkSide-50 (DS-50) \cite{DarkSide-50:2023fcw}, XENONnT \cite{XENON:2023cxc},  LUX-ZEPLIN (LZ) \cite{LZ:2022lsv}, XLZD \cite{Aalbers:2022dzr} and we have projected this on ($\lmd\,,\ms$) plane. As indeed, one can see, some parts of the parameter space are already ruled out by the existing experimental bounds from DS-50 (purple-shaded), XENONnT (magenta-shaded), and LZ (brown-shaded). Reheating, however, provides wider allowed parameter space for WIMP-like dark matter, which may be possible to detect in the future in experiments like XLZD within the gray region. For the FIMP-like DM, however, due to extremely weak coupling, all the parameter spaces are allowed. Note that the indirect DM detection constraint in Higgs portal models is superseded by those from direct detection experiments and the LHC. Therefore, we do not consider them in our analysis.

\section{QCD Axion as a DM}{\label{sc7}}    
In the earlier section, we discussed both thermal and non-thermal DM production of DM via both gravitational and Higgs portal interaction. In this section, we discuss the impact of our new reheating scenario on another well-known non-thermal production DM, which is identified as QCD axions \cite{Preskill:1982cy, Abbott:1982af,Dine:1982ah,Arias:2012az}. We primarily focus on exploring the parameter space where axions are produced during reheating through the misalignment mechanism \cite{Marsh:2015xka,DiLuzio:2020wdo}, accounting for the entire observed abundance of dark matter. Before diving into this topic, let’s briefly review the axion model, touching upon the key concepts and relevant parameters to be considered. The Lagrangian density for the axion field can be written as \cite{GrillidiCortona:2015jxo}
\begin{equation}{\label{axl}}
    \mathcal{L}_{\rm\mathfrak{a}}=\frac{1}{2}\partial^\mu\mathfrak{a}\,\partial_{\rm\nu}\mathfrak{a}-\Tilde{m}^2_{\rm\mathfrak a}(T)\left[1-\cos{\frac{\mathfrak a}{f_{\rm\mathfrak a}}}\right]\,,
\end{equation}
where $\ak$ is the axion field, $f_{\rm\mathfrak a}$ denotes the decay constant and $\Tilde{m}_{\rm\mathfrak a}(T)$ be the temperature-dependent
mass of the axion, which can be written as \cite{Borsanyi:2016ksw,Arias:2021rer,Arias:2022qjt}
\be
\label{attractorpotential}
\Tilde{m}_{\rm\mathfrak a}(T) \;=\;m_{\rm\mathfrak a}
\begin{cases}
(T_{\rm qcd}/T)^4\,&~\mbox{for}~~T\geq T_{\rm qcd}\,,\\
1\,&~\mbox{for}~~T\leq T_{\rm qcd}\,,\\
\end{cases}
\ee
where the QCD phase transition temperature $\tqc=150$ MeV and $m_{\rm\mathfrak a}$ identified as a zero-temperature axion mass which is given by
\be
m_{\mathfrak a}\simeq5.7\times10^{-6}\left(\frac{10^{12}\, \rm GeV}{f_{\mathfrak a}}\right)\,\rm eV
\ee
Using the Lagrangian \ref{axl}, one can find the following EoM for the zero mode axion field,
\be{\label{axeq}}
\ddot\theta+3H\dot\theta+\Tilde{m}^2_{\rm\ak}(T)\sin{\theta}=0\,,
\ee
where $\theta\equiv\ak/f_{\rm\ak}$. 
When the temperature is significantly higher than the QCD transition temperature ($T\geq\tqc$), the Hubble parameter is much larger than the axion's mass. In this regime, the axion remains static, essentially frozen in place. Axions start to oscillate when the temperature $T=\tos$ defined by $3H(\tos)\equiv\Tilde{m}_{\rm\ak}(\tos)$ \cite{kolb}. Assuming a standard radiation Universe, the corresponding
oscillation temperature can be calculated as
\be
\tos\simeq\;\begin{cases}
    \left(\frac{1}{\pi}\sqrt{10/g_\star(\tos)}\,m_{\rm \ak}\,\mp\right)^{1/2}&\tos\leq\tqc\,,\\
    \left(\frac{1}{\pi}\sqrt{10/g_\star(\tos)}\,m_{\rm \ak}\,\mp\tqc^4\right)^{1/6}&\tos\geq\tqc\\
\end{cases}
\ee
Below $\tos$, the axion behaves like a non-relativistic particle. Under the assumption that both the axion number density and the Standard Model (SM) entropy are conserved, the energy density of these non-relativistic axions at present, denoted by $\rho_\ak$:
\be\label{rhat0}
\rha(T_0)=\rha(\tos)\frac{\ma}{\mat(\tos)}\frac{s(T_0)}{s(\tos)}\,,
\ee
where $\rha(\tos)\simeq\frac{1}{2}\ma^2(\tos)f_\ak^2\,\theta_i^2$, $\theta_i$ be the initial misalignment angle. Using Eq.\ref{rhat0}, the axion abundance can be written as
\be
\begin{aligned}
\Omega_\ak h^2&\equiv\frac{\rha(T_0)}{\rho_{\rm c}/h^2}\\
&\simeq\left(\frac{\theta_i}{1.0}\right)^2\;\begin{cases}
   0.003 \left(\frac{\ma}{5.6\times10^{-6}\,\mbox{eV}}\right)^{-\frac{3}{2}}&\ma\leq\ma^{\rm qcd}\,,\\
     0.09\left(\frac{\ma}{5.6\times10^{-6}\,\mbox{eV}}\right)^{-\frac{7}{6}}&\ma\geq\ma^{\rm qcd}\,,\\
\end{cases}
\end{aligned}
\ee
with critical energy density $\rho_c=1.05\times10^{-5}\,h^2\,\rm {GeV/cm^3}$,$\,s(T_0)\simeq2.69\times10^3\,\rm cm^{-3}$ is the entropy density at today and $\ma^{\rm qcd}\simeq4.8\times10^{-11}$ eV. To match the observed DM abundance, the initial misalignment angle $\theta_{\rm i}$ could be tuned. Assuming an initial misalignment angle $\theta_{\rm i}\simeq 1.0$, the above equation indicates that an axion mass around $\ma\sim\mathcal{O}(10^{-5})$ eV and the decay constant $f_\ak\sim\mathcal{O}(10^{12})$ GeV is required to match the observed relic abundance.

 The initial misalignment angle is deeply connected to the cosmological history of the axion. If the Peccei-Quinn (PQ) symmetry is broken before or during inflation (and not restored afterward), the initial angle $\theta_i$ will be uniform across the observable universe, and the angle is randomly chosen from the range \([-\pi, \pi]\). For $\theta_{\rm i}\sim\mathcal{O}(1)$, $\ma$ should be in the range of $\mu$eV to avoid overproducing the DM abundance. Smaller axion masses (or equivalently, larger values of the decay constant $f_\ak$) become viable if small-tuned values of $\theta_i$ are considered.  This is often referred to as the anthropic axion window \cite{Wantz:2009it,Hertzberg:2008wr}. In this pre-inflationary scenario, large values of $f_\ak$ are constrained by isocurvature perturbations that impose $f_\ak\lesssim10^{16}$ GeV \cite{Arias:2021rer}. On the other hand,  if the PQ symmetry breaks
 after inflation, i.e, in a post-inflationary scenario, the initial value of $\theta_i$ may be different at different points in space, since we are interested in mode with zero momentum, therefore, we can safely use Eq. \ref{axeq}. When it breaks after the inflationary epoch, topological defects such as strings and domain walls could potentially contribute to axion production, affecting the mass range predictions \cite{Arias:2021rer,Gorghetto:2018ocs,Kawasaki:2014sqa,Hiramatsu:2010yu}. However, due to the persistent uncertainties regarding the extent of these contributions, our analysis has been limited to axion production through the misalignment mechanism. In this work, we primarily explore the impact of \(\nu\)GRe on the misalignment mechanism. Depending on the decay scale \(f_a\), both pre-inflationary ($f_\ak\geq\hend$) and post-inflationary ($f_\ak<\hend$) scenarios can emerge.

Now, we consider the situation where axion oscillation begins during the reheating phase, i.e., $\tos>\tre$. For such a case, the axion energy density at the present epoch is prescribed as \cite{Xu:2023lxw,Barman:2023icn,Grin:2007yg,Visinelli:2009kt,Arias:2021rer,Bernal:2021yyb}
\be\label{rhat}
\rha(T_0)=\rha(\tos)\frac{\ma}{\mat(\tos)}\frac{s(T_0)}{s(\tos)}\frac{s^\prime(\tos)}{s^\prime(\tre)}\,,
\ee
where the last factor takes into consideration the dilution of axion energy density due to entropy injection from the energy transfer between the inflaton/RHN and the thermal bath, and it can be written as
\be
\frac{s^\prime(\tos)}{s^\prime(\tre)}=\frac{g_{\star s}(\tos)}{g_{\star s}(\tre)}\left(\frac{\tos}{\tre}\right)^{\frac{3\mathcal{K}-3}{\mathcal{K}}}\,.
\ee
Here we utilize the relation $T=\tre(a/\are)^{-\mathcal{K}}$, where $\mathcal{K}=1\,(3/8)$ for the neutrino heating (neutrino dominating) case. $\tos$ can be different for different background reheating history. Now, we need to work out $\tos$, and it has two different possibilities: \\
\textbf{Case-I :}$\,\mathbf{\tos>\tqc}$\\
Let us suppose $w$ is the EoS of the dominant component during reheating, so $H$ can be written as
\begin{equation}
    H(a)=H_{\rm re}\left(\frac{a}{\are}\right)^{-3(1+w)/2}~~\left[H_{\rm re}=\frac{\pi}{3}\sqrt{\frac{g_\star(\tre)}{10}}\frac{\tre^2}{\mp}\right]
\end{equation}
Using the condition of oscillation, one can then obtain the expression for oscillation temperature as 
\begin{equation}
  \tos=\tre\left(\frac{1}{\pi}\sqrt{\frac{10}{g_\star(\tre)}}\frac{\ma\,\mp\,\tqc^4}{\tre^6}\right)^{\frac{2\mathcal{K}}{8\mathcal{K}+3(1+w)}}\,.
\end{equation}
Depending on the hierarchy between $\tre$ and $\tos$, we have two possibilities: $\tre <\tqc < T_{\rm osc}$, and that leads following inequality 
\be\label{tosc1}
\tqc^{\frac{3(1+w)}{3(1+w)-4\mathcal{K}}}\left(\frac{1}{\pi}\sqrt{\frac{10}{g_\star(\tre)}}\,\ma\,\mp\right)^{\frac{2\mathcal{K}}{4\mathcal{K}-3(1+w)}}<\tre<\tqc\,.
\ee
\begin{figure}[h]
\centering
\includegraphics[width=\columnwidth]{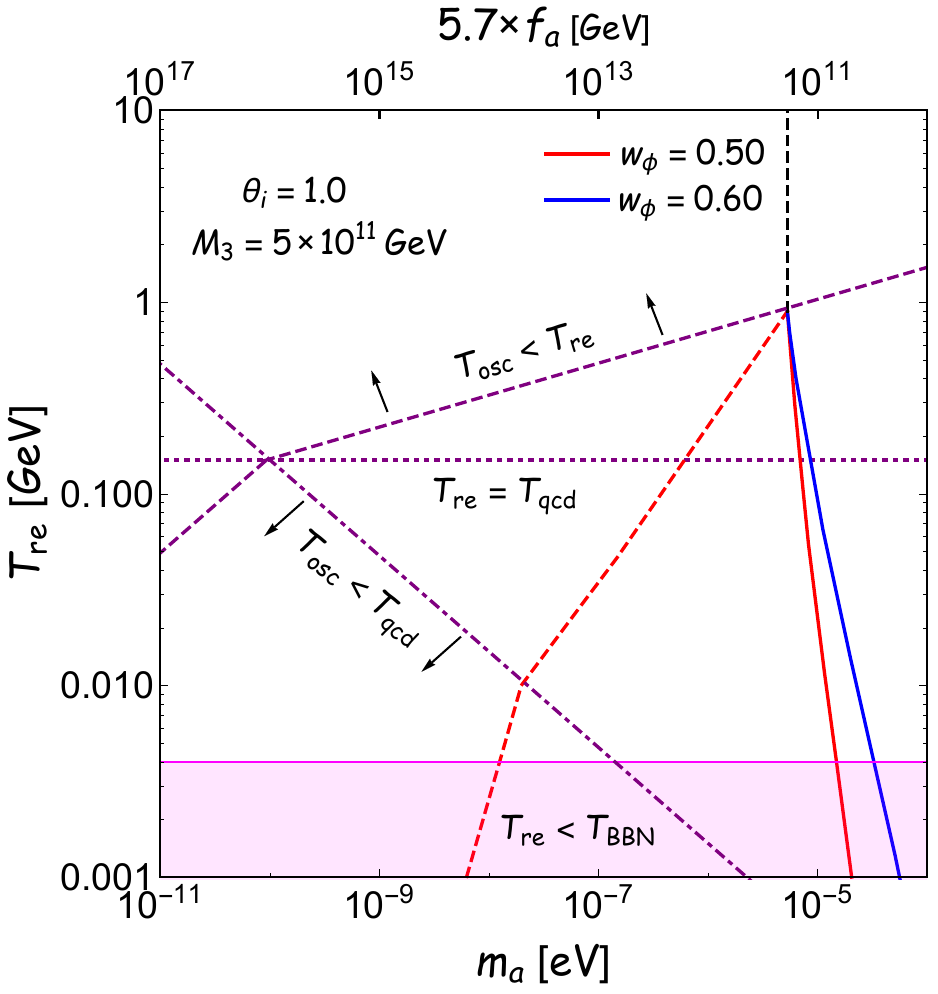}
\caption{The required $\tre$ as a function $\ma$ or ($f_{\rm\mathfrak a}$) from the observed DM abundance for $\wphi=0.50$ (red) and $\wphi=0.60$ (blue) with $\theta_i=1.0$. The purple dashed line corresponds to $\tre=\tos$, which separates  the parameter space into regions of
$\tre>\tos$ and $\tre<\tos$, above and below it, respectively.  The purple dot-dashed line corresponds to $\tos=\tqc$ (only for neutrino domination, i.e $w=0.0$), which separates  the parameter space into regions of
$\tos>\tqc$ and $\tos<\tqc$, above and below it, respectively. The horizontal purple dotted line indicates
$\tre=\tqc$. The magenta-shaded
regions are ruled out from $\tre<T_{\rm BBN}$. The vertical black dashed line lies above the $\tos=\tre$ lines, which indicates that oscillations occur after reheating, i.e., radiation-domination epoch and the corresponding axion mass $\ma\simeq6\times10^{-6}$ eV.}
\label{axp1}
\end{figure}
 Depending on the axion mass, we, therefore, obtained a range of reheating temperature $\tre$ for which the oscillation temperature $\tos$ follows the condition $\tre <\tqc < T_{\rm osc}$. For example for neutrino dominating case ($w=0\,,\mathcal{K}=3/8$) if we assume $\ma=10^{-7}$ eV, the reheating temperature should be constrained within $0.004<\tre<0.15$ GeV. However, given the axion mass, the DM abundance fixes the $\tre=0.05\,,\tos=0.3$ GeV and $\tre$ is  $<\tos=0.3$ GeV as can be decoded from the Fig.\ref{axp1}

The second possibility $\tqc<\tre<\tos$ leads to 
\be
\tqc<\tre<\left(\frac{1}{\pi}\sqrt{\frac{10}{g_\star(\tre)}}\,\ma\mp\right)^{1/6}\,.
\ee
 Again, using the same combination of ($w\,,\mathcal{K}\,,\ma$), as above the bounds on reheating temperature turns out to be  $0.15<\tre<2$ GeV. However, to satisfy the DM abundance, $\tre$ is required to be $0.05$ GeV, which is smaller than $\tqc =0.15 $ GeV. As a result, this inequality is not satisfied for $\ma=10^{-7}$ eV.

\textbf{Case-II :}$\,\mathbf{\tos<\tqc}$
\\
This is the case when the axion mass remains constant as expressed in Eq.\ref{attractorpotential}. The analytical expressions of $\tos$ for this case turn out as, 
\be\label{tosc2}
\tos=\tre\left(\frac{1}{\pi}\sqrt{\frac{10}{g_\star(\tre)}}\frac{\ma\mp}{\tre^2}\right)^{\frac{2\mathcal{K}}{3(1+w)}}
\ee
Similar to the earlier case, using the self-consistency condition $\tre<\tos$ and $\tre<\tqc$ , one can arrive at the following inequalities which the reheating temperature must satisfy,
\be
\begin{aligned}
&\tre<\tqc^{\frac{3(1+w)}{3(1+w)-4\mathcal{K}}}\left(\frac{1}{\pi}\sqrt{\frac{10}{g_\star(\tre)}}\,\ma\mp\right)^{\frac{2\mathcal{K}}{4\mathcal{K}-3(1+w)}}\,,\\
&\tre<\left(\frac{1}{\pi}\sqrt{\frac{10}{g_\star(\tre)}}\,\ma\mp\right)^{1/2}\,.
\end{aligned}
\ee
This is a constraint on $\tre$ where the oscillation temperature $\tos$ will satisfy the both conditions above. 
Particular for neutrino dominating case ($w=0\,,\mathcal{K}=3/8$) and assuming $\ma=1.5\times10^{-8}$ eV, observed DM abundance can be satisfied only for $\tre=6\times10^{-3}$ GeV and $\tos=0.095$ GeV, and this satisfies the above conditions which gives $\tre\leq0.01$ GeV (see Fig.\ref{axp1} for depiction).

Utilizing Eq.\ref{tosc1} and \ref{tosc2} in Eq.\ref{rhat}, we find the following expressions of the relic abundance,
\begin{equation}
    \begin{aligned}
        \Omega_\ak h^2&\simeq7.9\times10^5\left(\frac{\theta_i}{1.0}\right)^2\frac{g_{\star s}(T_0)}{g_{\star s}(\tre)}\\
       &~~~~ \begin{cases}
         \left(\frac{\tqc}{\tre}\right)^{\frac{3w-1}{1+w}} \left(\frac{\ma}{93\,\mbox{peV}}\right)^{-\frac{2}{1+w}}&\ma\leq\ma^{qcd}\,,\\
             \left(\frac{\tqc}{\tre}\right)^{7+6\mathcal{K}^\prime} \left(\frac{\ma}{93\,\mbox{peV}}\right)^{\mathcal{K}^\prime}&\ma\geq\ma^{qcd}\,,\\
        \end{cases}
    \end{aligned}
\end{equation}
where $\mathcal{K}^\prime=-2(4\mathcal{K}+3)/(8\mathcal{K}+3w+3)$.
\begin{figure*}
\centering
\includegraphics[width=16cm,height=7cm]{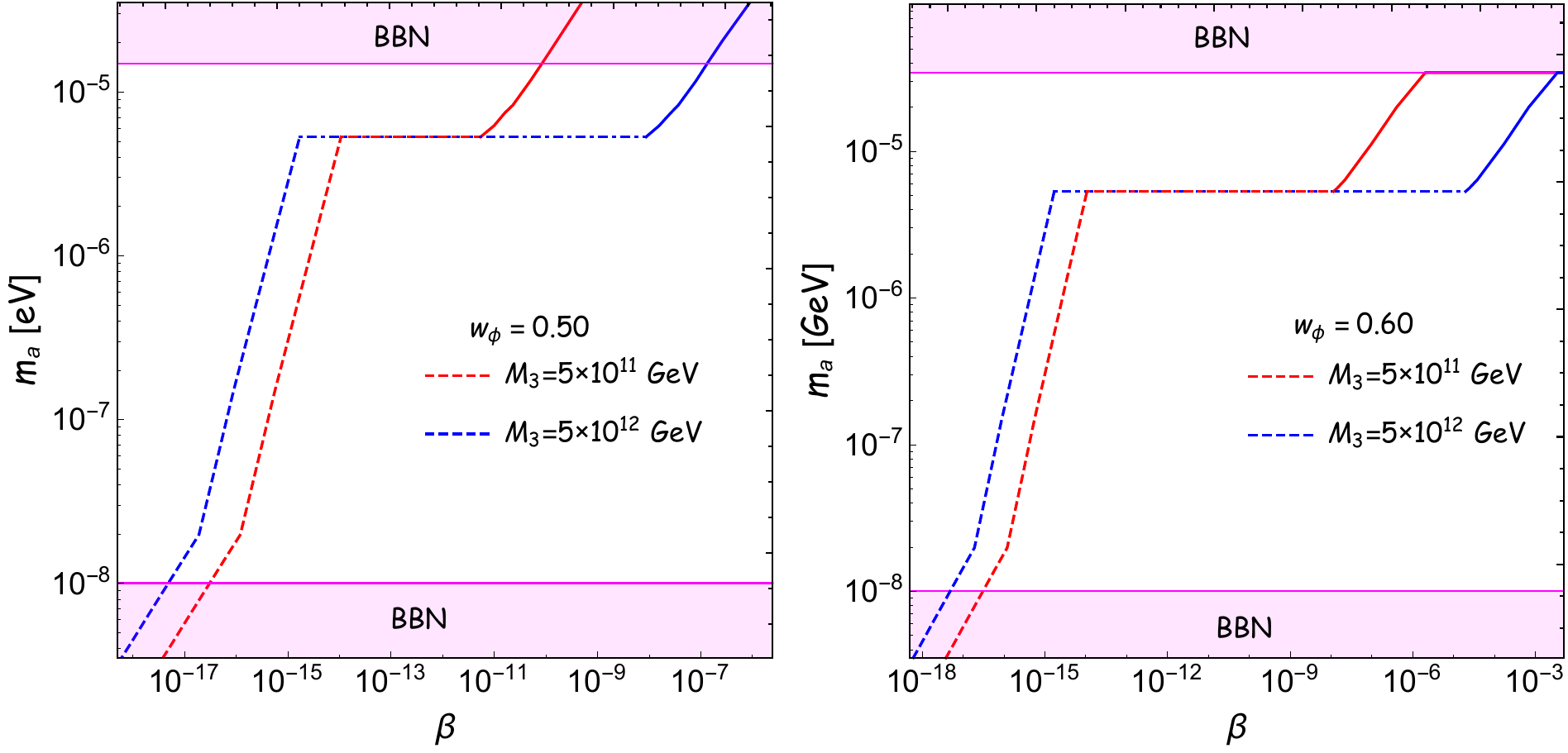}
\caption{The required $\ma$ as a function $\beta$ from the observed DM abundance with $\theta_i=1.0$ for two different choice $M_3:\{5\times10^{11}\,,2\times10^{13}\}$ GeV. The dashed (solid) line represents that misalignment occurs during the neutrino-dominating (heating) phase. The dot-dashed line represents that misalignment occurs after reheating, i.e., $\tos<\tre$. The magenta-shaded regions correspond to the regions where the required $\tre$ (for satisfying the current relic ) is less $T_{\rm BBN}$.  }
\label{axp2}
\end{figure*}

We have summarized our results in Fig.\ref{axp1}, which is consistent with the current axion DM abundance for  EoS, $\wphi=0.5\,(\mbox{red line})$ and $\wphi=0.6\,(\mbox{blue line})$. The dashed line represents the neutrino dominating, and the solid line corresponds to the neutrino heating scenario. The purple dashed line corresponds to $\tre=\tos$, which separates  the parameter space into regions of
$\tre>\tos$  and $\tre<\tos$ above and below the constant line respectively. The purple dot-dashed line corresponds to $\tos=\tqc$ (only for neutrino domination, i.e $w=0.0$), which separates  the parameter space into regions of
$\tos>\tqc$ and $\tos<\tqc$, above and below it, respectively. The horizontal purple dotted line indicates
$\tre=\tqc$. The magenta-shaded
regions are ruled out from $\tre<T_{\rm BBN}$. The vertical black dashed line lies above the $\tos=\tre$ lines, which indicates that oscillations occur after reheating, i.e., radiation-domination epoch and the corresponding axion mass $\ma\simeq6\times10^{-6}$ eV. Such tight bound has been shown to alter once the oscillation occurs during reheating \cite{Xu:2023lxw} due to additional entropy injection. The present paper deals with the effect of gravitational neutrino reheating on the axion parameter space. The final axion abundance depends on the background EoS as well as the evolution of the bath temperature (if $\tos>\tqc$). For the neutrino dominating case, the oscillation occurs during the phase when neutrino dominates the background, i.e. $w=0$, therefore $\Omega_{\mathfrak a}h^2$ turns out to be independent of inflaton EoS $\wphi$. For both $\wphi=0.50$ and $0.60$, the prediction is represented by the red dashed line, and the axion mass should lie within  $10^{-8}\,\mbox{eV}\leq\ma\leq6\times10^{-6}\,\mbox{eV}$. On the other hand, for neutrino heating case, $\Omega_{\mathfrak a}h^2$ depends on the inflaton EoS $\wphi$, therefore the prediction of $\ma\,\mbox{or}\,f_{\rm\mathfrak a}$ is different for different $\wphi$, and are represented by solid blue and red lines in Fig.\ref{axp1}. The allowed mass window for $\wphi=0.5\,(0.6)$, however, is approximately within $6\times10^{-6}\lesssim \ma\lesssim 2(4)\times10^{-5}$ eV.

For both reheating scenarios, as $\tre$ increases, each line converges toward the black dashed line. This is because, for larger $\tre$, the misalignment tends to occur after reheating so that the reheating dynamics do not
play any role. We found an upper limit of $\tre=1$ GeV for both reheating cases that ensures misalignment occurs during the reheating phase. We have further checked that for any values $\wphi\geq0.7$, the lower limit of $\tre$ for the neutrino heating case always remains larger than $1$ GeV, and hence mass bounds coincide with the black dashed line. Conversely, the neutrino reheating scenario turns out to be consistent with the axion DM scenario for $1/3\leq\wphi\leq0.7$ and very low $\tre \lesssim 1 \rm GeV$. 

Given the axion abundance the parameter space ($\tre\,,\ma$) is found to be nearly independent of $M_3$ values. As the neutrino coupling $\beta$ along with its mass $M_i$ control the value of $\tre$, we found an interesting correlation among seemingly disconnected DM and neutrino parameters ($\ma\,,\beta\,, M_3$) as depicted in  Fig.\ref{axp2}. The slanted lines correspond to neutrino dominating (dashed) and neutrino heating (solid) cases for two sample values of $M_3$. $\beta$ independent horizontal line corresponding to $\ma\simeq6\times10^{-6}$ eV which is the prediction of standard misalignment. To this end, we again reiterate that the new axion mass windows due to reheating emerge for the cases only when the reheating temperature is low $\lesssim1$ GeV.

\begin{figure}[t]
\centering
\includegraphics[width=\columnwidth]{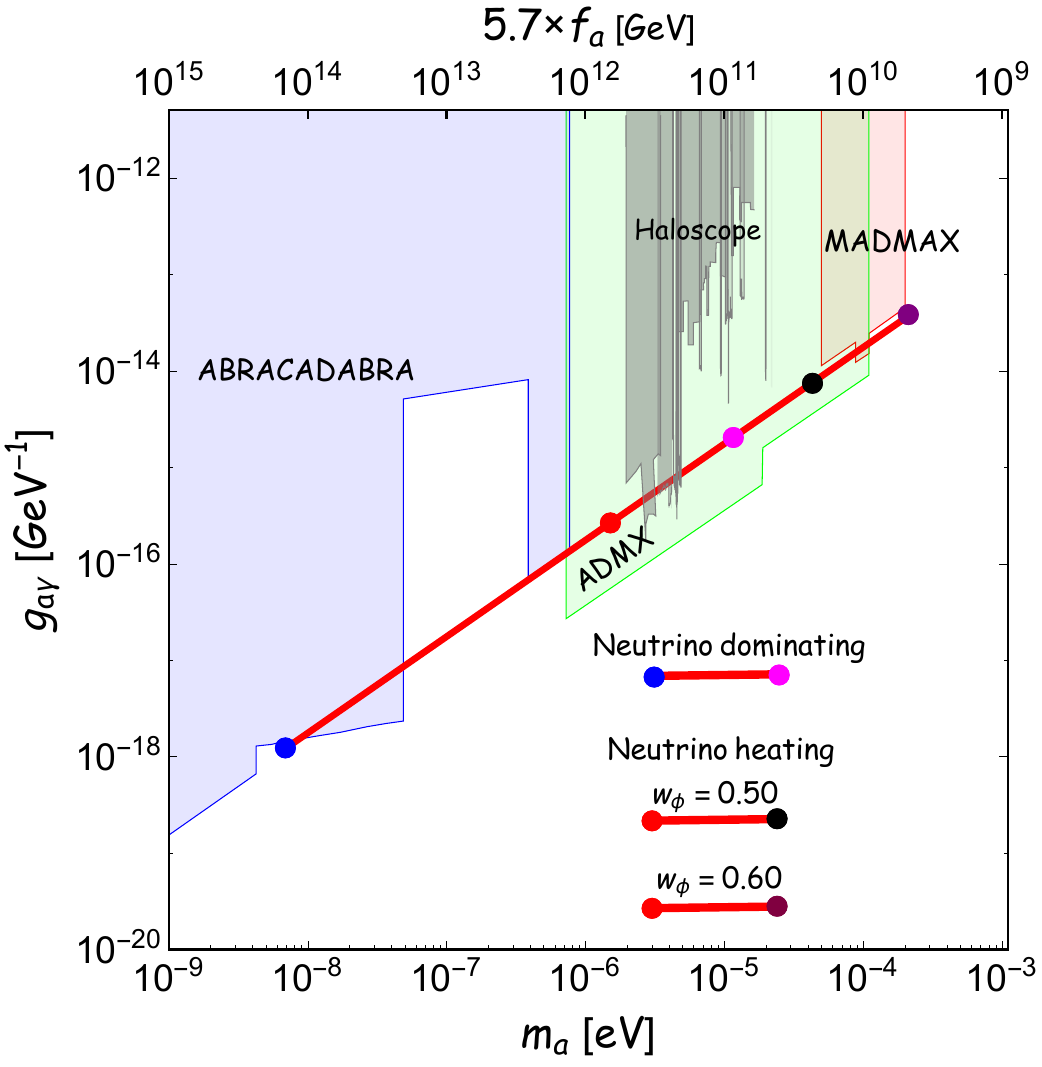}
\caption{The red lines correspond to the required axion-photon coupling $g_{\rm\mathfrak{a\gamma}}$ as a function of $\ma$ from the observed DM relic abundance for QCD axion by considering the $\nu$GRe. The segments between blue and magenta, as well as red and purple dots, correspond to the parameter space for neutrino dominating and neutrino heating cases, respectively.  The gray shaded areas
correspond to excluded parameter space from cosmology, astrophysics, and laboratory experiments, and different color-shaded areas show prospects in sensitivity from various experiments.}
\label{axexp}
\end{figure}
\subsection{Experimental constraints on axion parameter space}
In the previous section, we analyzed the relic-satisfied parameter space of axion characterized by its mass $\ma$ and the decay constant $f_{\rm\mathfrak a}$. However, such a parameter space can be further constrained from the present and proposed axion search facilities. The interaction between axions and two photons is one of the most commonly used channels in experiments and observations to detect the signature of axions. The interaction Lagrangian has taken the following form \cite{Marsh:2015xka,DiLuzio:2020wdo}
\be
\mathcal{L}_{\mathfrak {a\gamma}}=-\frac{1}{4}g_{\mathfrak {a\gamma}}\mathfrak a F_{\rm\mu\nu}\tilde F^{\mu\nu}\,,
\ee
where the coupling constant $g_{\mathfrak {a\gamma}}$ is related to the decay constant $f_\mathfrak{a}$ as \cite{Graham:2015ouw}
\be
\begin{aligned}
g_{\mathfrak {a\gamma}}&\simeq10^{-13}\,\mbox{GeV}^{-1}\left(\frac{10^{10}\,\mbox{GeV}}{f_{\mathfrak a}}\right)\\
&\simeq1.8\times10^{-15}\,\mbox{GeV}^{-1}\left(\frac{\ma}{10^{-5}\,\mbox{eV}}\right)
\end{aligned}
\ee

In Fig.\ref{axexp}, we show the parameter space ($g_{\mathfrak a\gamma},\ma$) or ($g_{\mathfrak a\gamma},f_{\rm\ak}$) for QCD axion with current and future experimental limit. The red line corresponds to the required axion-photon coupling $g_{\rm\mathfrak{a\gamma}}$ as a function of $\ma$ from the observed DM relic abundance for QCD axion by considering the effects of  $\nu$GRe. For finding the parameter space, we have taken $\wphi=0.50$ and $0.60$ with initial angle in the range $\theta_i\in(1/2,\pi/\sqrt{3})$. The segments between blue and magenta, as well as red and purple dots, correspond to the parameter space for neutrino dominating and neutrino heating cases, respectively. So, viable parameter space for $w_\phi=0.50\,(0.60)$ is $7\times10^{-9}\,\leq\ma\leq5\times10^{-5}\,(3\times10^{-4})\,$eV. Further, some portions of the viable parameter space are already constrained by the Haloscope experiments like ADMX \cite{ADMX:2018gho,ADMX:2019uok,Crisosto:2019fcj,ADMX:2021nhd,ADMX:2021mio} and CAPP \cite{Lee:2020cfj,CAPP:2020utb,Kim:2022hmg,Jeong:2020cwz}, as shown by gray-shaded regions. We have also shown the parameter space with future projections, such as MADMAX \cite{Beurthey:2020yuq}, which is shown by red-shaded, ADMX \cite{Stern:2016bbw} (green), and broadband axion-search experiment ABRACADABRA \cite{Ouellet:2018beu} (blue). All these experiments are capable of constraining the parameter space in further.

\section{Conclusions}{\label{sc8}}
After the inflation, the occurrence of reheating is still an unsettled issue. Primarily because observing such phenomena through cosmological experiments, even in the near future, is believed to be difficult. After realizing few existing indirect observational constraints such as BBN temperature $(T_{\rm BBN} \sim 4~ \mbox{MeV})$ \cite{Hannestad:2004px,Kawasaki:2000en}, effective number relativistic degrees of freedom $\Delta N_{\rm eff}\sim 2.99\pm0.34$ at $95\%$ CL \cite{Cyburt:2015mya,Planck:2018vyg}, dark matter abundance $(\Omega_{\rm DM} h^2 \sim 0.12)$ \cite{Planck:2018vyg}, even the CMB anisotropy parametrized by scalar spectral index $n_{\rm s} = 0.9649 \pm 0.0042$ at $68\%$ CL \cite{Planck:2018vyg}, are shown to have underlying connection with the reheating dynamics, a spate of research activity has emerged in the recent time \cite{Maity:2016uyn,Maity:2018dgy,Maity:2018exj}. Subsequently, several attempts have been made to understand this phase from the more fundamental point of view, invoking as much minimal parameters/ingredients as possible \cite{Lozanov:2017hjm,Lozanov:2016hid}. Purely gravitational interaction can lead to successful reheating is one such interesting example that has recently gained significant interest in the context of gravitational reheating \cite{Haque:2022kez}, minimal preheating \cite{Brandenberger:2023zpx},
reheating through non-minimal gravitational coupling \cite{Clery:2022wib}, and recently proposed gravitational neutrino reheating ($\nu$GRe) \cite{Haque:2023zhb} to name a few. 
 
In this paper we firstly discussed in great details this neutrino reheating scenario. Thanks to the well known Type-I seesaw model which is the most minimal extension of the standard model to simultaneously explain the active neutrino mass $\Delta m_i \sim 0.05\,\mbox{eV} $ and baryon asymmetry of the universe $Y_B \sim 10^{-10}$. The intriguing feature of this Type-I seesaw model is that apart from accounting all the aforementioned observable, it can further reheat the universe with following two distinct possibilities;

{\bf a) Neutrino dominating: $\beta\leq\beta_\nu^c$ ::} For this case, neutrino appears to be dominating at some intermediate scale factor $a_\nu$ during reheating. After this, it is the neutrino decay that naturally dominates the radiation production and sets the reheating temperature $(\tre)$, and is found to be completely insensitive to the inflation parameter $T_{\rm re}\propto\beta \,M_3^{1/2}$ (see dashed lines in Fig.\ref{pre}). $\tre$ is observed to reach its maximum at  $\beta=\beta_{\nu}^c$. For $w_\phi = 9/11$, the maximum temperature turned out to be as high as $10^{6}$ GeV, which is much larger than the temperature predicted in pure GRe case $\tre \sim 10^3$ GeV.  We found $\beta$ could be as low as $10^{-18}$ for minimum possible reheating temperature $T_{\rm BBN}$.
Because of the $\nu_{\rm R}^3$ behaving as matter, the PGW spectrum acquires a spectral break at $k=k_\nu$, where neutrino domination starts during the reheating phase. For $\kre < k < k_\nu$, the primordial GW spectrum behaves as $\Omega^{\rm k}_{\rm GW}h^2 \propto k^{-2}$, and for $\kend < k < k_\nu$, the spectrum behaves as $\Omega^{\rm k}_{\rm GW}h^2 \propto k^{-\frac{(2-6\,w_\phi)}{(1+3\,w_\phi)}}$ which is as expected inflaton parameter dependent.

{\bf b) Neutrino heating:$\beta^c_\nu\leq\beta\leq\beta_\phi^c$::}.
If we further increase $\beta>\beta^c_\nu$, even though neutrinos fail to dominate over inflaton, they still control the reheating through their decay. The reheating temperature is shown to behave as $T_{\rm re} \propto \beta^{\frac{-1}{3w_\phi-1}}
  M_{\rm 3}^{\frac{7+9w_\phi}{4(3w_\phi-1)}}$ acquires a negative power in $\beta$, and hence $\tre$ decrease with $\beta$ unlike the previouc case. For $w_{\rm\phi}\leq0.60$, the maximum allowed value of $\beta$ will be determined from the BBN bound of $T_{\rm re}\simeq 4$MeV, again for $w_{\rm\phi}>0.60$, the maximum value of $\beta$ is $\beta^{\rm c}_{\rm GW}$.

$\nu$GRe can Successfully reheat the universe if the inflaton EoS $\wphi$ lie within $1/3\leq\wphi\leq1.0$ and $\nu^3_{\rm R}$ mass $10^{10}\leq M_3\leq2\times10^{13}$ GeV. As a byproduct, we also obtain the non-vanishing lowest active neutrino mass due to non-zero $\beta$ requiring to have successful reheating. However, if one takes into account the right amount of baryon asymmetry at the present day, the reheating equation of state is further constrained to be $0.5\lesssim w_\phi\lesssim1.0$ in consistent with BBN and PLANCK. 

The effect of non-standard cosmology has gained significant interest in DM phenomenology.  In the second half of this paper, we explore the implications of $\nu$GRe for the production of DM. Firstly, we have considered a real scalar singlet DM, which interacts with the SM via the Higgs portal interaction and also with inflaton through universal gravitational interaction. Depending on the strength of the portal coupling $\lambda$, the production of the DM particle $S$ can occur either through thermal or non-thermal processes. For the well-known freeze-out mechanism for the weakly interacting massive particle(WIMP), the portal coupling typically has to be much larger compared to the freeze-in mechanism for feebly interacting massive particles (FIMP). The required value of the portal coupling $\lmd$ to achieve the observed DM abundance is plotted as a function of DM mass $m_{\rm s}$ in Fig.\ref{SSD2} and \ref{SSD3} for both FIMP (top plot) and WIMP (bottom plot) type DM. For the FIMPs production,  the viable mass range turned out to be within $10^2\,\mbox{eV}\leq m_s\leq10\,
  \mbox{GeV}$. Note that any values  lower than the bound of $100\, \mbox{eV}$ FIMP mass will give the under abundance today \cite{Haque:2023yra} due to extremely weak production. On the other, mass $>10\,
  \mbox{GeV}$, gives over-abundance due to strong gravitational production. Additionally the perturbativity conditions restrict the mass of the WIMP DM, within $1\,\leq m_s\leq4\times10^5 \,\mbox{GeV}$ for $\wphi=1/2$ and $1\,\leq m_s\leq 5\times10^7 \,\mbox{GeV}$ for $\wphi =9/11$. Finally, we have shown the impact of direct detection experiments on our parameter space. Reheating, however, provides wider allowed parameter space for WIMP-like DM which may be possible to detect in the future in experiments like XLZD. For the FIMP-like DM, however,
 due to extremely weak coupling, all the parameter spaces are allowed.

 Finally we investigated the impact of $\nu$GRe on the production of QCD axions as a candidate of DM via the vacuum misalignment mechanism. We analyzed the relic-satisfied parameter space (see Fig.\ref{axp1}) of axion characterized by its mass $\ma$ and the decay constant $f_{\rm\mathfrak a}$. Furthermore, we analyzed the experimental constraints by comparing this parameter space with both current and upcoming axion experiments (see Fig.\ref{axexp}). We found that the current ADMX and CAPP experiments are capable of excluding small portions of the expanded parameter space, particularly within the mass range $2\times10^{-6}\,\leq\ma\leq5\times10^{-6}\,$eV. The remaining regions of the parameter space could be explored in future experiments.

\section{acknowledgements}
RM would like to thank the Ministry of Human Resource Development, Government of India (GoI),
for financial assistance. MRH wishes to acknowledge support from the Science and Engineering Research Board (SERB), Government
of India (GoI), for the SERB National Post-Doctoral fellowship, File Number: PDF/2022/002988. DM wishes
to acknowledge support from the Science and Engineering Research Board (SERB), Department of Science and
Technology (DST), Government of India (GoI), through the Core Research Grant CRG/2020/003664. RM would like to thank Nayan Das and Disha Bandyopadhyay for the helpful discussions. DM and RM also thank the Gravity and High Energy Physics groups at IIT Guwahati for illuminating discussions. We thank the anonymous referee for his critical review and comments which helped our paper greatly.
\onecolumngrid
\appendix
\section{Construction of $y_{ij}$ using Casas-Ibarra Parametrization}\label{CIP} 
When the neutral component of the SM Higgs doublet acquires a VEV leading to the spontaneous breaking of the SM gauge symmetry, neutrinos in the SM acquires a Dirac mass matrix $m_{\rm D} =y\,v $. Where $v=174$ GeV is the vacuum expectation value of the SM Higgs. Using this Dirac mass $m_{\rm D }$ together with the bare Majorana mass $M$, the light neutrino mass matrix can be written as $m_\nu\simeq m^{\rm T}_{\rm D} M^{-1}m_{\rm D} $.
The left-handed neutrino masses are obtained by diagonalizing the mass matrix $m_{\rm\nu}$ with the Pontecorvo-Maki-Nakagawa-Sakata (PMNS) unitary matrix $U$ \cite{pdg} as $m^{\rm d}_{\rm\nu}=\mbox{diag}(m_1,m_2,m_3)=U^{\rm T}m_{\rm\nu}U$.
By using this well-known Casas-Ibara (CI) parametrization \cite{Casas:2001sr}, Yukawa coupling $y$ can be written as
 \begin{equation}
     y=\frac{1}{v}U\sqrt{m^{\rm d}_{\rm\nu}}\mathbb R^{T}\sqrt{M} .
 \end{equation}
 We choose the orthogonal matrix $\mathbb R$ as
 \begin{equation}
 \mathbb R=
 \begin{pmatrix}
0 & \cos{z} & \sin{z}\\
0 & -\sin{z} & \cos{z}\\
1 & 0 & 0
\end{pmatrix},
\end{equation}
where $z = a+ib$ is a complex angle. The diagonal light neutrino mass matrix $m^{\rm d}_{\rm\nu}$
is calculable
using the best-fit values of solar and atmospheric mass obtained from the latest neutrino
oscillation data \cite{Esteban:2018azc}. The elements of Yukawa coupling matrix $y$ for a specific value
of $z$ can be obtained for different choices of the heavy neutrino masses. 
For example, with $M_{1}=3\times10^{12}$ GeV,$M_2=5\times10^{13}$ GeV, $\{a,b\}=\{2.5,1.4\}$, using  normal hierarchy, $m_3=0.05$ eV,$m_2=0.008$ eV and taking $M_3$, $m_1$ are the free parameter, we obtain
\begin{equation}
    y=
\begin{pmatrix}
-0.043-0.013\,i &\quad  -0.043+0.130\,i & \quad0.821\,\sqrt{m_1\,M_3}/v\\
0.032-0.087\,i &\quad  -0.315-0.0889\,i & \quad\quad(0.301-0.062\,i)\,\sqrt{m_1\,M_3}/v\\
0.079-0.044\,i &\quad  -0.161-0.220\,i & \quad(0.477-0.055\,i)\,\sqrt{m_1\,M_3}/v
\end{pmatrix} .
\end{equation}
In our present scenario, we need the
 third RHN to be the long-lived one, and its decay width should be very small.  By tuning the lightest
 active neutrino mass ($m_1$), we can tune the $\nu^3_{\rm R}$ decay width as $y_{\rm i3}\propto \sqrt{m_1M_3}/v$. To find a small but non-vanishing decay width, the lightest
 active neutrino mass should be non-vanishing. For example, taking $M_3=5\times10^{11}$ GeV, $m_1=2.5\times10^{-17}$ eV,  
 which produces the desired CP asymmetry, the coupling matrix assumes,
 \begin{equation}
    y^\dagger y=
\begin{pmatrix}
0.019 &\quad  -0.0051+0.0616i & \quad6.46\times10^{-37}+4.84\times10^{-27} i\ \\
-0.005-0.061 i\ &\quad  0.201 & \quad-1.292\times10^{-26}i\\
6.462\times10^{-27}-4.846\times10^{-27}i &\quad  1.292\times10^{-26}i & \quad4.131\times10^{-19}
\end{pmatrix}
\end{equation}
Note that the component of $(y^\dagger y)_{33}$, which is identified with $\beta$ parameter, is very small compared to $(y^\dagger y)_{11}\,,(y^\dagger y)_{22}$. Therefore, the decay width of the third RHN is suppressed compared to the other two RHNs, and this turned out to be crucial for successful neutrino reheating.

Similarly, we can construct the Yukawa coupling $y$ using the inverted hierarchy (IH). The primary distinction lies in the choice of the orthogonal matrix, where the form of orthogonal matrix $\mathbb R$ is,
\begin{equation}
 \mathbb R=
 \begin{pmatrix}
 \cos{z} & \sin{z} & 0\\
 -\sin{z} & \cos{z} &0\\
 0 & 0 & 1
\end{pmatrix}\,.
\end{equation}
For IH, the neutrino masses are given by $m_2=0.05$ eV, $m_1=0.0497$ eV, and $m_3$ is treated as a free parameter. Using these values, we obtain the following Yukawa coupling,
\begin{equation}
    y=
\begin{pmatrix}
-0.044-0.112\,i &\quad  -0.402+0.125\,i & \quad(0.111-0.100\,i)\,\sqrt{m_3\,M_3}/v\\
0.073-0.032\,i &\quad  -0.120-0.199\,i & \quad\quad0.738\,\sqrt{m_3\,M_3}/v\\
-0.107+0.024\,i &\quad  0.086+0.300\,i & \quad0.657\,\sqrt{m_3\,M_3}/v
\end{pmatrix} .
\end{equation}
Therefore, for IH, $y_{\rm i3}\propto \sqrt{m_3\,M_3}/v$ i.e, $\beta^2\propto m_3.$
\section{RG Effects in Neutrino Yukawa coupling}
The one-loop renormalization group evolution (RGE) equations of the relevant parameters
associated with the Seesaw-I model are :\\
The gauge couplings:
\begin{equation}
    16\,\pi^2\frac{dg_1 }{d\ln{\mu}}=\frac{41}{10}g_i^3,\,\,\,16\,\pi^2\frac{dg_2 }{d\ln{\mu}}=-\frac{19}{6}g_2^3,\,\,\,16\,\pi^2\frac{d\,g_3 }{d\ln\mu}=-7\,g_i^3
\end{equation}
 The Higgs quadratic and quartic couplings:
\[16\,\pi^2\frac{d\mu^2_H}{d\ln\mu}
 = \left(-\frac{9}{10} g_1^2 - \frac{9}{2} g_2^2 + 12\lambda + 2\,T\right)\,\mu^2_H
\]
\[16\,\pi^2\frac{d\lambda_H}{d\ln\mu}
 = \left(-\frac{9}{5} g_1^2  - 9 g_2^2 + 24 \lambda_H + 4T\right)\,\lambda_H+ \frac{27}{200} g_1^4 + \frac{9}{20} g_1^2 g_2^2 + \frac{9}{8} g_2^4-2T1
\]
 The Yukawa couplings:
\[16\,\pi^2\frac{dY_u}{d\ln\mu}=
 \left( -\frac{17}{20} g_1^2 - \frac{9}{4} g_2^2 - 8 g_3^2 + T \right) Y_u+ \frac{3}{2}(Y_u Y_u^\dagger Y_u)  - \frac{3}{2}(Y_u Y_d^\dagger Y_d)
\]
\[
16\,\pi^2\frac{dY_d}{d\ln\mu}= -\frac{1}{4} \left( g_1^2 + 9 g_2^2 + 32 g_3^2 -4\,T \right) Y_d + \frac{3}{2} (Y_d Y_d^\dagger Y_d) - \frac{3}{2}(Y_d Y_u^\dagger Y_u)
\]
\[
16\,\pi^2\frac{dY_e}{d\ln\mu}= \left( -\frac{9}{4} g_1^2 - \frac{9}{4} g_2^2 +T \right) Y_e + \frac{3}{2} (Y_e Y_e^\dagger Y_e)-\frac{3}{2} (Y_e Y_\nu^\dagger Y_\nu)
\]
\[
16\,\pi^2\frac{dY_\nu}{d\ln\mu}= \left( -\frac{9}{20} g_1^2 - \frac{9}{4} g_2^2 +T \right) Y_\nu + \frac{3}{2} (Y_\nu Y_\nu^\dagger Y_\nu)-\frac{3}{2} (Y_\nu Y_e^\dagger Y_e)
\]
The Majorana mass:
\[16\,\pi^2\frac{dM_\nu}{d\ln\mu}=
2  M_v (Y_v Y_v^\dagger)^T + Y_v Y_v^\dagger M_v
\]
where $T=  \text{Tr}\left(3Y_d Y_d^\dagger + 3Y_u Y_u^\dagger+ Y_e Y_e^\dagger+Y_\nu Y_\nu^\dagger\right)$ and $T1=\text{Tr}\left(3(Y_d Y_d^\dagger)^2 + 3(Y_u Y_u^\dagger)^2 + (Y_e Y_e^\dagger)^2+(Y_\nu Y_\nu^\dagger)^2\right)$
SARAH provides with a numerical solution to the derived renormalization group equations. In the figure \ref{runn}, we have shown ratio of running of $|y_{33}|$ and $|y_{22}|$ with the energy scale $\mu$. Initially, $\frac{|y_{33}|}{|y_{22}|}=9.4\times10^{-10}$, after taking running, $\frac{|y_{33}|}{|y_{22}|}\simeq9.9\times10^{-10}$ at $\mu=4$ MeV.
\begin{figure}[t]
\centering
\includegraphics[width=8cm,height=6cm]{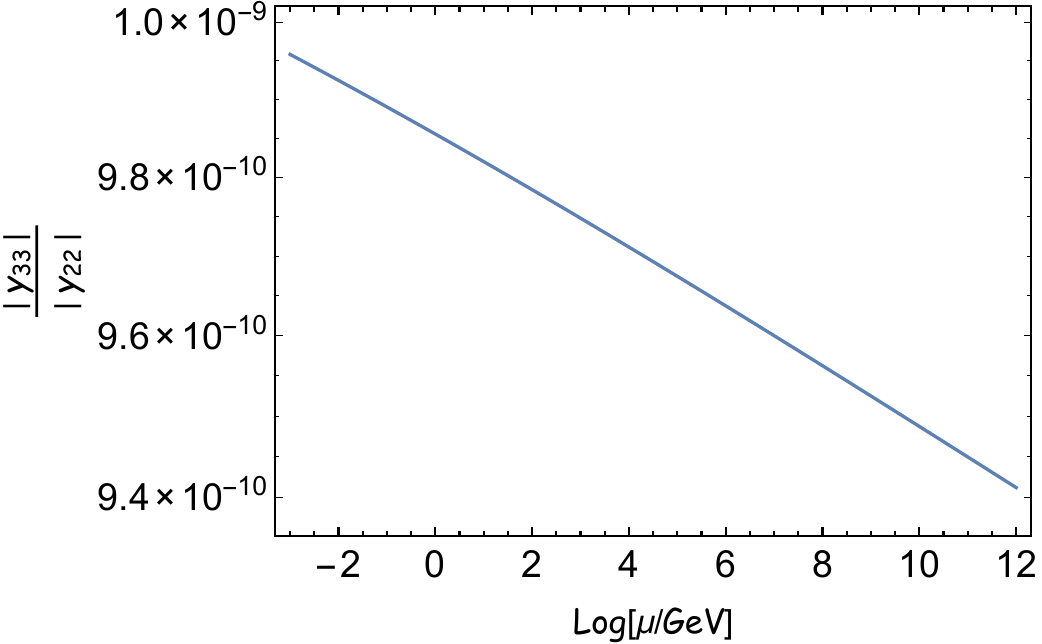}
\caption{Evolution of  $\frac{|y_{33}|}{|y_{22}|}$ with energy scale $\mu=T^{\rm max}_{\rm rad}$ to $\mu=T_{\rm BBN}=4$ MeV.}
\label{runn}
\end{figure}
\section{Listed the DM annihilation cross-section}\label{sigmav}
The thermally averaged cross-section $\langle\sigma v\rangle$ for any $2$-$2$ annihilation process can be calculated using the standard formula \cite{Gondolo}
\begin{equation}
    \langle\sigma v\rangle=\frac{g^2}{n^2_{\rm eq}}\frac{T}{32\pi^4}\int^{\infty}_{4m^2_{\rm s}}\sqrt{s}(s-4m^2_{\rm s})\sigma(s)K_1\left(\frac{\sqrt{s}}{T}\right)ds
\end{equation}
\begin{figure}[t]
\centering
\includegraphics[width=12cm,height=9cm]{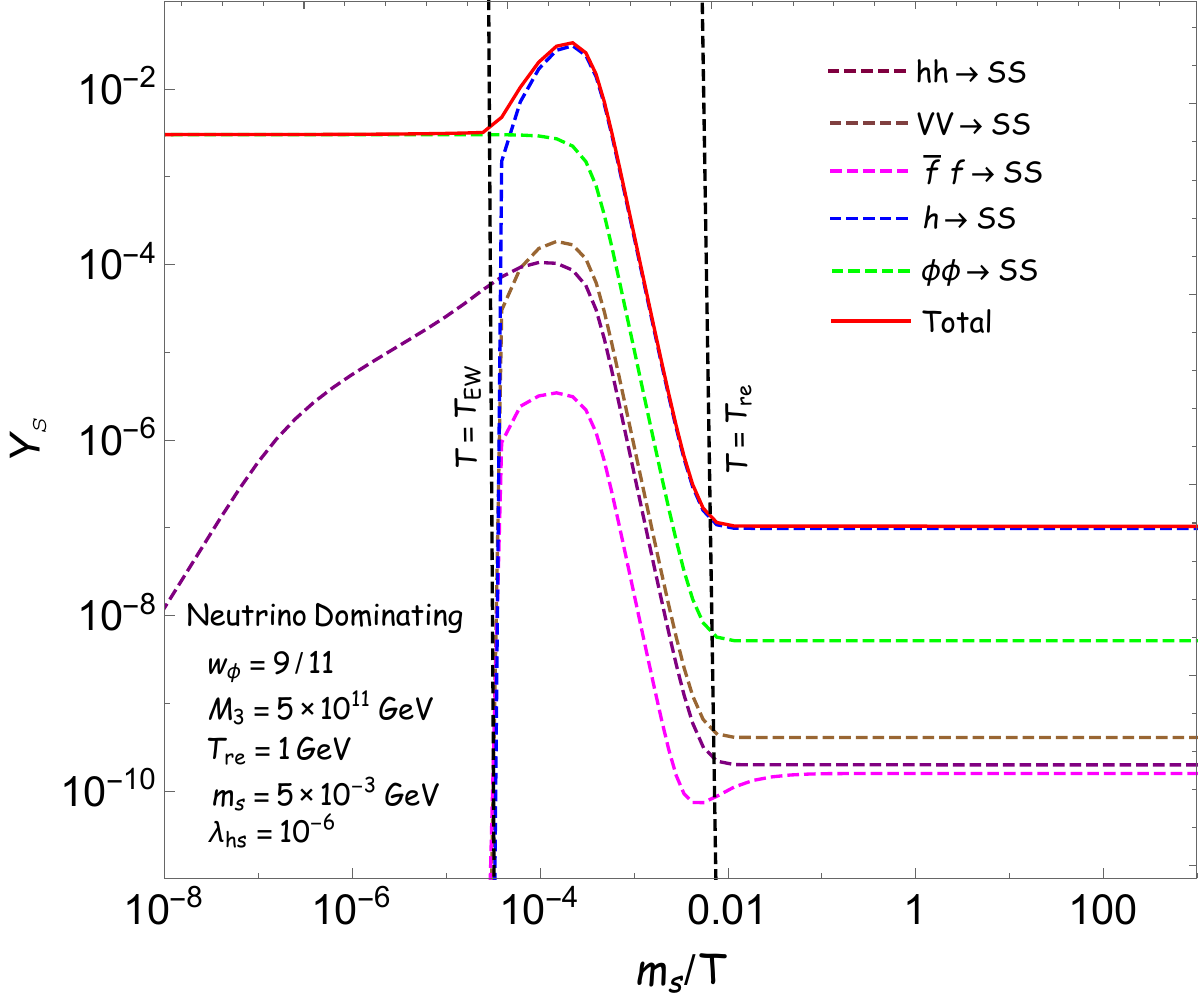}
\caption{ Evolution of DM yield (for freeze-in) as a function of inverse temperature with individual contributions (explained in inset) from different production processes; final yield
 (solid red line) corresponds to the correct DM relic.}
\label{yieldall}
\end{figure}
where $g=1$ is the internal
degrees of freedom for the scalar dark matter. $\sigma(s)$ and $s$ are the annihilation cross section and the squared center-of-mass energy of the DM particle respectively. The annihilation cross-section $\sigma(s)$ for different annihilation channels have the following forms:
\begin{equation}
   \sigma_{SS\rightarrow \bar ff}=\sum_{f}n_f\frac{\lambda^2_{\rm hs}m^2_{\rm f}}{2\pi\,s}\frac{\sqrt{s-4m^2_f}}{(s-m^2_{\rm h})^2+\Gamma_{\rm h}^2m_{\rm h}^2}\sqrt{\frac{s-4m^2_{\rm f}}{s-4\ms^2}},
\end{equation}
\begin{equation}
   \sigma_{SS\rightarrow VV}=\sum_{V}n_V\frac{\lambda^2_{\rm hs}}{4\pi s}\frac{(s^2-4s\,m^2_V+12m^4_V)}{(s-m^2_{\rm h})^2+\Gamma_{\rm h}^2m_{\rm h}^2}\sqrt{\frac{s-4m^2_{\rm V}}{s-4\ms^2}},
\end{equation}
\begin{equation}
\begin{aligned}
   \sigma_{SS\rightarrow hh}=&\frac{\lambda^2_{\rm hs}}{8\pi\,s}\sqrt{\frac{s-4m^2_{\rm f}}{s-4\ms^2}}\left[\zeta^2-\frac{16\zeta\,\lambda\, v^2}{s-2\mh^2}\frac{1}{2\xi}\log\left(\frac{1+\xi}{1-\xi}\right)+\frac{32\lambda^2 v^4}{(s-2\mh^2)^2}\left(\frac{1}{1-\xi^2}+\frac{1}{2\xi}\log\left(\frac{1+\xi}{1-\xi}\right)\right)\right],
   \end{aligned}
\end{equation}
where $\zeta=\left(\frac{s+2\mh^2}{s-\mh^2}\right)\,,\xi=\frac{\sqrt{(s-4\ms^2)(s-4\mh^2)}}{s-2\mh^2}$. $(\bar f)\,f$ denotes SM (anti-) fermions and  $n_f=3$ for quarks and $n_f=1$ for leptons. $V=W^{\pm}\,,Z$ denotes the SM gauge bosons and $n_W=1\,,n_Z=1/2$. 

To facilitate our understanding further into the DM evolution and identify the contribution of different production channels, in Fig.\ref{yieldall}, we have shown the evolution of total DM yield $Y_{\rm s}=n_{\rm s}/s$ for the freeze-in mechanism as a function of inverse temperature represented by solid red line. Along with this we further plotted individual contributions from different production processes in dotted lines. The total
 contribution, as denoted by the solid red line, reaches an asymptotic value, resulting the correct relic abundance. Before EWSB, the DM is produced only through Higgs annihilation and gravitational interaction. If the DM mass is chosen to be below the Higgs mass, as for the case in the figure, after EWSB, the DM can also be produced from the Higgs decay $h\rightarrow SS$ and the other scattering channels namely $VV\rightarrow SS,\bar ff\rightarrow SS$. However, Higgs decay is the dominant production channel as expected and indeed can be seen from the figure. Since $\tre<\mh/2$,  most of the DM is produced during reheating, and the yield $Y_{\rm s}$ suffers from a high entropy dilution if considering individual decay channel.

\twocolumngrid
\hspace{0.5cm}
 
\end{document}